\documentclass[11pt]{article}
\usepackage[final]{yuhao-template}
\usepackage{graphicx}

\title{Tight Analysis of Randomized Truthful Scheduling: \\An $(m+1)/2$ Lower Bound}
\author{%
  Wenqian Wang\thanks{Shanghai Jiao Tong University, China. Email: \texttt{wangwenqian@sjtu.edu.cn}.}
  \and
  Chenyang Xu\thanks{East China Normal University, China. Email: \texttt{cyxu@sei.ecnu.edu.cn}.}
  \and
  Yuhao Zhang\thanks{Shanghai Jiao Tong University, China. Email: \texttt{zhang\_yuhao@sjtu.edu.cn}.}
}
\date{}

\begin{document}

\maketitle

\begin{abstract}
The Nisan--Ronen conjecture, a central problem in algorithmic game theory, asserts that no deterministic truthful mechanism for makespan minimization on $m$ unrelated machines can achieve an approximation ratio better than $m$. Christodoulou, Koutsoupias, and Kov{\'a}cs resolved it in 2023, but the optimal approximation ratio achievable by randomized truthful mechanisms remained open. Recently, Anastasiadis, Christodoulou, Koutsoupias, Kov{\'a}cs, and Schecker extended techniques from the deterministic proof to establish a $(3/2-\sqrt{2})(m-1)$ lower bound for universally truthful mechanisms with discrete support and gave a universally truthful mechanism with approximation ratio $m/2+o(m)$.
For the broader class of truthful-in-expectation (TIE) mechanisms, however, the gap remained large: the best known upper bound was $(m+5)/2$, while the lower bound was only $2-1/m$. We prove a lower bound of $(m+1)/2$, determining the optimal TIE approximation ratio up to an additive constant as $m/2+\Theta(1)$. Our result also improves the leading constant in the lower bound for universally truthful mechanisms from $3/2-\sqrt{2} \approx 0.0858$ to $1/2$. Together with their $m/2+o(m)$ upper bound, our result determines the optimal approximation ratio for universally truthful mechanisms up to lower-order terms.

Our approach differs fundamentally from the deterministic lower-bound techniques and applies even to truthful fractional mechanisms, with approximation measured against the integral optimum. Its key ingredient is \emph{job allocation interaction}, which quantifies how a job's allocation depends on other jobs and provides a bridge between truthfulness and approximation lower bounds. We show that any truthful fractional mechanism with a finite approximation guarantee admits a replacement with essentially the same guarantee and no interactions of order at least $m$. We then prove that every fractional allocation rule with such bounded interaction has approximation ratio at least $(m+1)/2$, yielding the desired lower bound.
\end{abstract}

\clearpage

\section{Introduction}

The Nisan--Ronen conjecture is a central problem in algorithmic game theory. In 1999, Noam Nisan and Amir Ronen~\cite{DBLP:conf/stoc/NisanR99} gave a deterministic truthful $m$-approximation mechanism for makespan minimization on $m$ unrelated machines. They conjectured that no deterministic truthful mechanism could achieve a better approximation ratio. In this model, $n$ jobs must be assigned to $m$ machines, each of which privately knows its processing time for every job. A mechanism maps the reported processing times to an allocation and payments, and each machine seeks to maximize its utility, defined as its payment minus its actual processing cost. Truthfulness requires truthful reporting to be a dominant strategy, while the scheduling objective is to minimize the makespan, the maximum load over all machines. The conjecture thus concerns the approximation loss imposed by truthfulness alone, regardless of computational power.

Resolving the conjecture took more than two decades of research. Early work established a sequence of constant lower bounds~\cite{DBLP:conf/soda/ChristodoulouKV07,DBLP:conf/mfcs/KoutsoupiasV07,DBLP:conf/sagt/GiannakopoulosH20,DBLP:journals/corr/abs-2007-04362} for general deterministic truthful mechanisms and tight lower bounds of $m$ for restricted classes, including strongly monotone mechanisms~\cite{DBLP:conf/soda/MualemS07} and anonymous mechanisms~\cite{DBLP:conf/sigecom/AshlagiDL09}, but a large gap remained for general mechanisms. A breakthrough came when Christodoulou, Koutsoupias, and Kov{\'a}cs proved a lower bound of $1+\sqrt{m-1}$~\cite{DBLP:conf/focs/0001KK21}, showing that the approximation loss must grow with the number of machines. In 2023, they established the tight bound of $m$~\cite{DBLP:conf/stoc/0001KK23}, finally resolving the conjecture.

The tight deterministic bound leaves open how much randomization can improve the approximation ratio. In the randomized setting, there are two standard notions of truthfulness: \emph{universal truthfulness} and \emph{truthfulness in expectation}. A universally truthful mechanism samples a deterministic truthful mechanism from a distribution independent of the reports. A truthful-in-expectation (TIE) mechanism requires truthful reporting to maximize each machine's expected utility for every fixed profile of reports from the other machines, where the expectation is over the mechanism's internal randomness. Universal truthfulness implies TIE, whereas TIE permits a broader class of mechanisms by imposing incentive constraints only in expectation. For randomized mechanisms, we measure approximation by comparing the expected makespan with the optimal integral makespan.

Recently, Anastasiadis, Christodoulou, Koutsoupias, Kov{\'a}cs, and Schecker~\cite{DBLP:journals/corr/abs-2609-12621} determined the optimal approximation ratio for universally truthful mechanisms with discrete support up to a constant factor. They established a linear lower bound of $(3/2-\sqrt{2})(m-1)$ and constructed a universally truthful mechanism with approximation ratio $m/2+o(m)$, improving on the previous guarantee of approximately $0.837m$~\cite{DBLP:conf/stacs/LuY08}. Their lower-bound proof extends the Box Theorem underlying the deterministic $m$ lower bound to deterministic mechanisms with arbitrary finite approximation guarantees and combines it with a hard input distribution through Yao's minimax principle~\cite{DBLP:conf/focs/Yao77}.

Before our work, a substantial gap remained between the best known bounds for general TIE mechanisms: Mu'alem and Schapira~\cite{DBLP:conf/soda/MualemS07} proved a lower bound of $2-1/m$, while Lu and Yu~\cite{DBLP:conf/wine/LuY08a} gave a $(m+5)/2$-approximation. The lower bound for universally truthful mechanisms did not directly resolve this gap, since a TIE mechanism need not admit a representation as a random mixture of deterministic truthful mechanisms. The central question was therefore whether requiring truthfulness only in expectation permitted an approximation ratio sublinear in the number of machines.

We establish a lower bound of $(m+1)/2$ for TIE scheduling, showing that a linear approximation barrier persists even under this weaker incentive requirement.

\begin{center}
\begingroup
\setlength{\fboxsep}{10pt}
\setlength{\fboxrule}{0.5pt}
\fbox{%
\begin{minipage}{\dimexpr0.90\linewidth-2\fboxsep-2\fboxrule\relax}
{\centering\textbf{Our Result}\par}\smallskip
\emph{No truthful-in-expectation randomized
mechanism for makespan minimization on $m$ unrelated machines
can achieve an approximation ratio better than $(m+1)/2$.}
\end{minipage}%
}
\endgroup
\end{center}

\begin{figure}[htbp]
    \centering
    \input{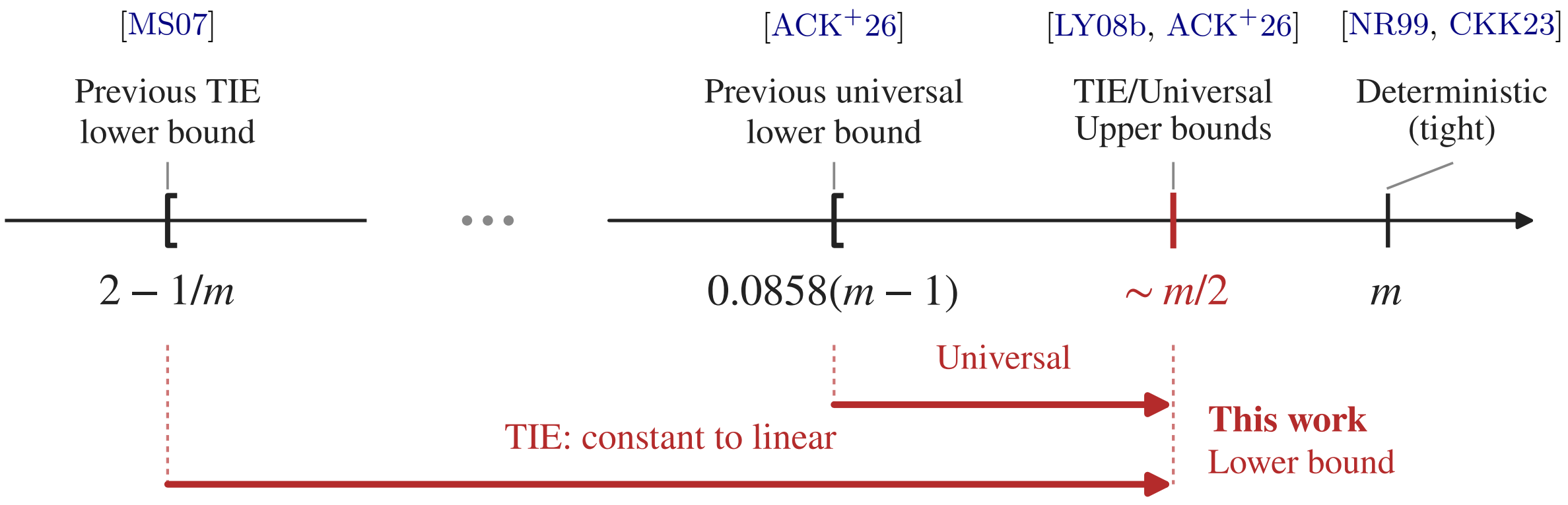}
    \caption{Red arrows show our lower-bound improvements. The mark $\sim m/2$ groups bounds whose ratios to $m/2$ tend to $1$ as $m\to\infty$. More precisely, our lower bound is $(m+1)/2$ for both classes, while the TIE and universal upper bounds are $(m+5)/2$~\cite{DBLP:conf/wine/LuY08a} and $m/2+o(m)$~\cite{DBLP:journals/corr/abs-2609-12621}, leaving gaps of $2$ and $o(m)$, respectively.}
    \label{fig:approximation-bounds}
\end{figure}

Our lower bound matches the $(m+5)/2$ upper bound~\cite{DBLP:conf/wine/LuY08a} for TIE scheduling up to an additive constant of $2$. Moreover, since every universally truthful mechanism is truthful in expectation, our lower bound applies to all universally truthful mechanisms. We thus improve the leading constant in the previous lower bound~\cite{DBLP:journals/corr/abs-2609-12621} from $3/2-\sqrt{2}\approx 0.0858$ to $1/2$ and remove the discrete support assumption. Together with the $m/2+o(m)$ upper bound from the same work, our result determines the optimal approximation ratio for universally truthful scheduling up to lower-order terms.
\cref{fig:approximation-bounds} summarizes these bounds.

\subsection{Technical Overview}\label{sec:tech}

This subsection outlines the main ideas behind our lower bound. Taking expectations of a TIE mechanism's allocations and payments yields a deterministic truthful fractional mechanism without worsening its approximation guarantee against the integral optimum. It therefore suffices to establish the desired lower bound on the worst-case ratio of fractional makespan to optimal integral makespan for every such mechanism.\footnote{By establishing the lower bound for truthful fractional mechanisms, we also match the $(m+1)/2$ upper bound of Christodoulou, Koutsoupias, and Kov{\'a}cs~\cite{DBLP:journals/talg/ChristodoulouKK10}, determining the exact optimal approximation ratio for truthful fractional scheduling.}

\paragraph{Quantifying job allocation interaction.} A key idea in our analysis is to quantify how a mechanism's allocation of a job to a machine changes as other jobs appear, and then use the resulting notion of \emph{job allocation interaction} as a bridge between truthfulness and approximation lower bounds.
More formally, we quantify job allocation interaction using a M\"obius decomposition~\cite{DBLP:journals/mor/GrabischMR00}, which separates the contributions of different groups of jobs. For each target job and machine, the decomposition associates an interaction coefficient with every subset of jobs containing that job. The size of the subset defines its interaction order. The first-order coefficient describes the job's allocation in isolation, while a second-order coefficient records the change caused by adding one other job. Higher-order coefficients capture the joint effects of larger groups after subtracting the contributions from smaller subsets. M\"obius inversion recovers each job's allocation by summing its interaction coefficients over subsets of the jobs present, yielding a multilinear polynomial representation in the indicators of job presence.

This notion connects the two main steps of our proof. We say that an allocation rule has interaction order at most $m-1$ if every coefficient in its M\"obius decomposition involving at least $m$ jobs is zero. First, we show that any deterministic truthful fractional mechanism with a finite approximation guarantee admits another such mechanism with essentially the same approximation guarantee and interaction order at most $m-1$, where $m$ is the number of machines. Second, for any fractional allocation rule with interaction order at most $m-1$, we construct a random instance for which the ratio of its expected makespan to the expected integral optimum is at least $(m+1)/2$, up to an arbitrarily small error. Together, these two steps establish the desired lower bound for deterministic truthful fractional mechanisms and hence for TIE mechanisms.

Previously, Christodoulou, Koutsoupias, and Kov{\'a}cs~\cite{DBLP:journals/talg/ChristodoulouKK10} proved a $(m+1)/2$ lower bound for task-independent fractional allocation rules, where each job's allocation depends only on its own processing-time vector. Our notion of interaction order generalizes task-independence, which corresponds to interaction order at most $1$. Our second step extends their lower bound to the broader class of allocation rules with interaction order at most $m-1$. Our first step connects this class to truthfulness by showing that every truthful fractional mechanism with a finite approximation guarantee admits a truthful replacement in this class with essentially the same guarantee.

\subsubsection{Using Truthfulness to Bound Interaction Order}

Truthfulness provides a key link between allocations and utilities: the fraction of a job assigned to a machine equals the negative partial derivative of that machine's utility with respect to its processing time for the job~\cite{DBLP:journals/geb/ArcherK14}. Because the M\"obius decomposition preserves this relation, we study allocation interactions through \emph{utility interactions}, defined by applying the same decomposition to utilities. For each job set $J$, the allocation interaction coefficients are the negative partial derivatives of the corresponding utility interaction coefficients. We then show that every deterministic truthful fractional mechanism with a finite approximation guarantee has a corresponding mechanism of the same type that essentially preserves this guarantee and has zero utility interaction coefficients for every job set $J$ with $|J|\ge m$. The derivative relation implies that the corresponding allocation interaction coefficients are also zero, giving an interaction order of at most $m-1$.

\paragraph{Laplace transforms of utility interactions.}
Consider a job set $J$ with $|J|\ge m$. Since each job in $J$ is fully allocated, its corresponding allocation interaction coefficients sum to zero across all $m$ machines. The derivative relation therefore yields one equation per job in which the corresponding utility derivatives also sum to zero. We apply the Laplace transform to the utility interactions and rewrite these equations in terms of their transforms. Each utility derivative contributes a multiple of the corresponding transformed interaction, together with an extra term involving the original utility interaction when the relevant processing time is zero. If these extra terms are zero, the equations form a homogeneous linear system with $|J|$ equations and $m$ unknowns, where the unknowns are the transformed utility interactions for the $m$ machines. This system provides the basis for proving that all utility interaction coefficients associated with $J$ are zero. To eliminate the extra terms, we need each utility interaction to be zero whenever the corresponding processing time is zero.

\paragraph{Zero boundary conditions via padding invariance.}
Although a truthful fractional mechanism need not satisfy the zero boundary conditions required above, we can always transform it into another truthful fractional mechanism that satisfies these conditions and preserves its approximation guarantee. We begin with a \emph{padding-appending operation}: append a \emph{padding job}, which has zero processing time on at least one machine, run the mechanism on the enlarged instance, and keep its allocations of the original jobs with suitably adjusted payments. Each operation preserves truthfulness and the approximation guarantee. The idea is to average over repeated padding operations so that further padding has progressively less effect. A common fixed-point argument~\cite{kakutani1938two} makes this intuition precise and yields a mechanism unchanged by every padding-appending operation. The resulting mechanism satisfies \emph{padding invariance}: adding any padding job leaves the allocations of existing jobs unchanged. Combined with the allocation--utility relation, this invariance ensures that a machine's utility interaction is zero whenever one of the jobs involved has zero processing time on that machine, giving precisely the zero boundary conditions needed above.

With the zero boundary conditions in place, the Laplace transform gives a homogeneous linear system. For $|J|\ge m$, a rank argument based on the density of matrices with full column rank~\cite{DBLP:books/daglib/0086372} shows that its unknowns have to be zero. Uniqueness of the Laplace transform and the derivative relation then yield the desired bound of $m-1$ on interaction order.

\subsubsection{A Lower Bound under Bounded Interaction}

Consider a fractional allocation rule with interaction order at most $m-1$. We construct a random star instance by assigning the roles of its center and $m-1$ leaves uniformly at random to the machines. Each edge represents $L$ jobs restricted to its two endpoints, and we add $L$ jobs restricted to the center. Each job has processing time $1/L$ on every allowed machine. We include each job independently with probability $\rho$ and compare two functions of $\rho$: the rule's expected center load $F(\rho)$, which lower-bounds its expected makespan, and the expected integral optimum $\operatorname{OPT}(\rho)$.

\paragraph{Singleton instances and initial slopes.}
We call an instance containing exactly one job a \emph{singleton instance}. For fixed $L$, nonempty instances contain a single job with probability tending to one as $\rho\downarrow0$. Splitting a singleton edge job equally between its endpoints gives a makespan of $1/(2L)$, half its integral optimum. In our center-load calculation, the same factor follows from role symmetry: either endpoint is equally likely to be the center, and exchanging their roles leaves the singleton instance unchanged. Each edge job therefore contributes $1/(2L)$ to the expected center load when it appears alone, while a center-only job contributes $1/L$. Since every singleton has integral optimum $1/L$, the initial slopes are $\operatorname{OPT}'(0)=m$ and $F'(0)=(m+1)/2$.

\paragraph{The gap between expected load and optimum.}
As $\rho$ increases, instances with more jobs become more likely, and an integral schedule can process the $m$ groups in parallel by assigning the center-only jobs to the center and each edge group to its leaf. This allows $\operatorname{OPT}(\rho)/\rho$ to fall from $m$ to a value close to $1$. For fixed $m$, increasing $L$ moves this transition arbitrarily close to the origin. Bounded interaction prevents $F(\rho)/\rho$ from following this rapid decrease for two reasons: (i) interaction order at most $m-1$ makes $F$ a polynomial of degree at most $m-1$, independently of $L$; and (ii) the construction ensures that $0\le F(\rho)\le m\rho$ throughout $[0,1]$. Thus, $F(\rho)/\rho$ is a polynomial of bounded degree with values in $[0,m]$, so its rate of decrease near the origin is bounded independently of $L$. Choosing $\rho$ sufficiently small and then $L$ sufficiently large keeps $F(\rho)/\rho$ arbitrarily close to $(m+1)/2$ while bringing $\operatorname{OPT}(\rho)/\rho$ arbitrarily close to $1$, yielding the approximation lower bound of $(m+1)/2$.

\subsection{Other Related Work}

\paragraph{Deterministic and randomized truthful scheduling.}
Early lower bounds for deterministic truthful mechanisms progressed from $2$ to $1+\sqrt{2}$~\cite{DBLP:conf/soda/ChristodoulouKV07}, then to $(3+\sqrt{5})/2$~\cite{DBLP:conf/mfcs/KoutsoupiasV07}, and subsequently to approximately $2.755$~\cite{DBLP:conf/sagt/GiannakopoulosH20} and arbitrarily close to $3$~\cite{DBLP:journals/corr/abs-2007-04362}. For randomized mechanisms, Nisan and Ronen~\cite{DBLP:conf/stoc/NisanR99} give a universally truthful $7/4$-approximation for two machines. Mu'alem and Schapira~\cite{DBLP:conf/soda/MualemS07} obtain a $0.875m$-approximation for general $m$, and Lu and Yu~\cite{DBLP:conf/stacs/LuY08} improve this guarantee to approximately $0.837m$. The two-machine case has also received further attention: Lu and Yu~\cite{DBLP:conf/stacs/LuY08} obtain a $1.6737$-approximation, while Chen, Du, and Zuluaga~\cite{DBLP:journals/mst/ChenDZ15} give a copula-based mechanism with approximation ratio at most $1.58606$.

\paragraph{Truthful scheduling in restricted settings.}
Several works study truthful scheduling under restrictions on processing times or machine eligibility. Lavi and Swamy~\cite{DBLP:journals/geb/LaviS09} use cycle monotonicity to design truthful mechanisms for domains with two publicly specified processing times per job, obtaining a deterministic $2$-approximation when these two values are common to all jobs. Auletta, Christodoulou, and Penna~\cite{DBLP:journals/mst/Auletta0P15} examine domains in which each machine's private information consists of a single bit and exhibit a separation between universal truthfulness and truthfulness in expectation. Christodoulou, Koutsoupias, and Kov{\'a}cs~\cite{DBLP:conf/icalp/0001KK21} study graph-restricted scheduling, where each job can be processed only by two prescribed machines, and establish upper and lower bounds for several graph classes, including stars and trees.

\paragraph{Related machines.}
Truthful scheduling has also received extensive attention on related machines, where the classical model assumes public job sizes and private machine speeds. Archer and Tardos~\cite{DBLP:conf/focs/ArcherT01} characterize truthful implementability through monotonicity and show that makespan minimization admits an optimal truthful mechanism without computational restrictions. Dhangwatnotai, Dobzinski, Dughmi, and Roughgarden~\cite{DBLP:conf/focs/DhangwatnotaiDDR08} give a truthful-in-expectation PTAS, while Christodoulou and Kov{\'a}cs~\cite{DBLP:conf/soda/ChristodoulouK10} give a deterministic truthful PTAS. Epstein, Levin, and van Stee~\cite{DBLP:conf/soda/EpsteinLS13} develop a unified framework for deterministic truthful approximation schemes covering makespan minimization and other load objectives. Online variants also consider strategic jobs with private sizes. With public machine speeds, Feldman, Fiat, and Roytman~\cite{DBLP:conf/sigecom/FeldmanFR17} obtain a constant-competitive dynamic posted-price mechanism that is truthful for jobs. Li, Li, and Wu~\cite{DBLP:journals/scheduling/LiLW23} study additional requirements on the distribution of workloads across machines and give an $O(\log m)$-competitive, job-truthful mechanism producing almost well-behaved schedules. More recently, Wang, Xu, and Zhang~\cite{WangXZ26} give an $O(\log m)$-competitive online mechanism on related machines that is truthful in expectation for both machines and jobs.

\section{Preliminaries}
\label{sec:preliminaries}
We study approximation lower bounds for truthful-in-expectation (TIE) randomized mechanisms for makespan minimization on unrelated machines. Our lower bound holds more generally for truthful fractional mechanisms, even when their makespan is compared with the optimal integral makespan. Every TIE randomized mechanism induces a truthful fractional mechanism by taking expected allocation shares and payments. This transformation preserves expected utilities and produces a fractional makespan no larger than the original expected makespan. We define the fractional model below and give the reduction at the end of this section.

\paragraph{Instances and allocations.}
An instance consists of $m\ge 2$ machines $[m]=\{1,\ldots,m\}$, $n$ jobs $[n]=\{1,\ldots,n\}$, and a processing-time matrix $p\in\mathbb{R}_{\ge 0}^{m\times n}$. The entry $p_{ij}$ is the processing time of job $j$ on machine $i$, and is also the cost incurred by that machine for processing the entire job. All processing times are finite, and zero processing times are allowed. Machine $i$ privately knows its entire row $p_i=(p_{ij})_{j\in[n]}$; we write $p_{-i}$ for the rows of the other machines. The job set is public, and each machine may report any vector in $\mathbb{R}_{\ge 0}^n$.

A fractional allocation is a matrix $x\in\mathbb{R}_{\ge 0}^{m\times n}$ satisfying
$$
\sum_{i=1}^m x_{ij}=1
\qquad\text{for every }j\in[n]~.
$$
Here $x_{ij}$ is the fraction of job $j$ assigned to machine $i$, and processing this fraction incurs cost $p_{ij}x_{ij}$. Let $\mathcal{X}([n])$ denote the set of feasible fractional allocations. The makespan of $x$ is the maximum load over all machines,
$$
C(p,x)=\max_{i\in[m]}\sum_{j\in[n]}p_{ij}x_{ij}~.
$$
An allocation is integral if $x_{ij}\in\{0,1\}$ for every $i,j$. Let $\mathcal{X}_{\mathrm{int}}([n])$ denote the set of feasible integral allocations. Throughout the paper, our benchmark is the optimal integral makespan,
$$
\operatorname{OPT}(p)=\min_{x\in\mathcal{X}_{\mathrm{int}}([n])}C(p,x)~.
$$
When adding or removing jobs, we also allow an arbitrary finite set of job labels $S\subseteq\mathbb{N}$. All definitions above extend to such a set by using $S$ in place of $[n]$ as the job index set.

\paragraph{Mechanisms and truthfulness.}
A mechanism family $M=(x^S,\pi^S)_S$ specifies, for every finite job set $S$, an allocation rule $x^S:\mathbb{R}_{\ge 0}^{m\times S}\to\mathcal{X}(S)$ and a payment rule $\pi^S:\mathbb{R}_{\ge 0}^{m\times S}\to\mathbb{R}^m$. The payment $\pi_i^S(p)$ is made to machine $i$. Payments are finite at every report, including reports with zero processing times. We write $x_i(p)=(x_{ij}(p))_{j\in S}$ for machine $i$'s allocation vector and use $p_i\cdot x_i=\sum_{j\in S}p_{ij}x_{ij}$ for its processing cost.

Utilities are quasilinear: each machine values its payment linearly and subtracts its processing cost. When machine $i$ has true processing times $p_i$ and reports $\hat p_i$, while the other machines report $p_{-i}$, its utility is $\pi_i(\hat p_i,p_{-i})-p_i\cdot x_i(\hat p_i,p_{-i})$. The family is \emph{truthful} if, for every finite job set, every machine $i$, and all $p_i,\hat p_i,p_{-i}$,
$$
\pi_i(p)-p_i\cdot x_i(p)\ge\pi_i(\hat p_i,p_{-i})-p_i\cdot x_i(\hat p_i,p_{-i})~.
$$
Thus truthful reporting maximizes each machine's utility for every fixed report of the other machines. Deviations may change the machine's entire processing-time vector.

\paragraph{Normalized utility.}
We define machine $i$'s normalized utility under truthful reporting by
$$
u_i(p)=\pi_i(p)-p_i\cdot x_i(p)-\pi_i(0,p_{-i})~,
$$
where $0$ denotes the all-zero row on the current job set. The normalization subtracts a term independent of machine $i$'s own report and therefore preserves its incentives. In particular, $u_i(0,p_{-i})=0$. We write $u_i^S$ when the job set needs to be explicit, and use $u_i^\varnothing=0$ for the empty job set.

\paragraph{Approximation guarantee.}
For fixed $m$, the family is \emph{$\Gamma$-approximate} if one finite constant $\Gamma$ satisfies
$$
C\bigl(p,x^S(p)\bigr)\le\Gamma\operatorname{OPT}(p)
$$
for every finite nonempty job set $S$ and every processing-time matrix $p$ with $\operatorname{OPT}(p)>0$. The ratio $\Gamma$ may depend on $m$, but not on the number of jobs or their processing times. The mechanism and its truthfulness requirement are defined on the full nonnegative domain even when the optimum is zero.

\paragraph{From randomized mechanisms to fractional mechanisms.}
Consider a randomized mechanism family. Fix a job set $[n]$. On report $p$, the mechanism outputs a random integral allocation $X(p)\in\mathcal{X}_{\mathrm{int}}([n])$ and payments $\Pi(p)$ with finite expectations. The family is truthful in expectation if truthful reporting maximizes each machine's expected utility for every fixed report of the other machines. Define a deterministic fractional mechanism by
$$
x_{ij}(p)=\mathbb{E}[X_{ij}(p)],
\qquad
\pi_i(p)=\mathbb{E}[\Pi_i(p)]~.
$$
Expectations are over the mechanism's internal randomness. Since every realized allocation is feasible, $x(p)$ is a feasible fractional allocation. Moreover, for every true row $p_i$, deviation $\hat p_i$, and report $p_{-i}$,
$$
\pi_i(\hat p_i,p_{-i})-p_i\cdot x_i(\hat p_i,p_{-i})
=\mathbb{E}\bigl[\Pi_i(\hat p_i,p_{-i})-p_i\cdot X_i(\hat p_i,p_{-i})\bigr]~.
$$
Thus the fractional mechanism preserves the randomized mechanism's expected utility under every report, and TIE implies truthfulness of the fractional mechanism.

For the objective, we have
$$
\begin{aligned}
C\bigl(p,x(p)\bigr)
&=\max_{i\in[m]}\mathbb{E}\left[\sum_{j\in[n]}p_{ij}X_{ij}(p)\right]\\
&\le\mathbb{E}\left[\max_{i\in[m]}\sum_{j\in[n]}p_{ij}X_{ij}(p)\right]\\
&=\mathbb{E}\bigl[C\bigl(p,X(p)\bigr)\bigr]~.
\end{aligned}
$$
Consequently, a $\Gamma$-approximation guarantee in expectation for the randomized mechanism, measured against $\operatorname{OPT}(p)$, gives the same $\Gamma$-approximation guarantee for the induced fractional mechanism. The argument allows arbitrary correlations among job assignments. Any approximation lower bound for truthful fractional mechanism families against this integral benchmark therefore also holds for TIE randomized mechanism families.

\section{Proof Framework}
\label{sec:proof-framework}

We now formalize the proof framework outlined in~\cref{sec:tech}. The key notion is \emph{job allocation interaction}, which links truthfulness to approximation lower bounds through two ingredients: a reduction from truthful fractional mechanisms to mechanisms of interaction order at most $m-1$, and a $(m+1)/2$ lower bound for every fractional allocation family with this bounded interaction order. The reduction preserves any finite approximation guarantee up to an arbitrarily small additive error per job, which uniform bounds on the hard instances allow us to absorb. We begin by defining job allocation interaction and its order, then state the two ingredients and combine them to prove the main theorem.


\subsection{Job Allocation Interaction}
\label{sec:job-interactions}

We apply the M\"obius decomposition to each job's allocation as a function of the set of jobs present. Consider an instance $p$ on $[n]$, and let $p_S$ denote the submatrix consisting of the columns in $S\subseteq[n]$. For each machine $i$ and job $j$, consider the set function that takes the value $x_{ij}^S(p_S)$ when $j\in S$ and zero otherwise. Here, $x^S(p_S)$ is computed by running the mechanism on the smaller instance $p_S$, so it need not agree with the corresponding columns of $x^{[n]}(p)$. This gives the following coefficients.

\begin{definition}[Job allocation interaction]
\label{def:job-interaction}
For a finite job set $J$, a job $j\in J$, and a machine $i\in[m]$, define the \emph{job allocation interaction coefficient} by
\[
G_{ij}^J(p_J)
=\sum_{\substack{S\subseteq J\\j\in S}}
(-1)^{|J|-|S|}x_{ij}^S(p_S).
\]
A mechanism has \emph{interaction order at most $d$} if, for any finite job set $J$ with $|J|>d$ and any nonnegative report $p_J$, we have $G_{ij}^J(p_J)=0$ for all $j\in J$ and $i\in[m]$.
\end{definition}

The coefficient $G_{ij}^J(p_J)$ captures the contribution of the jobs in $J$ to job $j$'s allocation on machine $i$, after subtracting the contributions associated with all proper subsets of $J$ containing $j$. Its order $|J|$ counts all jobs in $J$, including the target job $j$.

\paragraph{Interpreting job allocation interaction.}
Job interactions are discrete analogues of higher-order mixed derivatives: we take successive finite differences as jobs are added. Fix a machine $i$ and focus on job $1$. Write $f(S)=x_{i1}^S(p_S)$ when $1\in S$, and $f(S)=0$ otherwise, with the processing times fixed throughout. For a background set $S$ not containing job $1$, let $D(S)=f(S\cup\{1\})-f(S)$ be the effect of adding job $1$ to that background. The interaction coefficients measure how this effect changes across backgrounds.


The first-order coefficient is $G_{i1}^{\{1\}}=D(\varnothing)=f(\{1\})$: it is the effect of adding job $1$ to the empty instance. The second-order coefficient is
$$
G_{i1}^{\{1,2\}}
=D(\{2\})-D(\varnothing)
=f(\{1,2\})-f(\{1\})~.
$$
It measures how the effect of adding job $1$ changes when job $2$ is already present.

The third-order coefficient takes one more difference:
$$
\begin{aligned}
G_{i1}^{\{1,2,3\}}
&=\bigl[D(\{2,3\})-D(\{2\})\bigr]
 -\bigl[D(\{3\})-D(\varnothing)\bigr]\\
&=f(\{1,2,3\})-f(\{1,2\})-f(\{1,3\})+f(\{1\})~.
\end{aligned}
$$
The first bracket measures how adding job $3$ changes the effect of job $1$ when job $2$ is present; the second measures the same change when job $2$ is absent. Thus, the third-order coefficient records how the second-order interaction changes in the presence of another job. The change from background $\{2\}$ to $\{2,3\}$ alone is only the first bracket; subtracting the corresponding change from $\varnothing$ to $\{3\}$ removes the interaction already present between jobs $1$ and $3$. Higher-order coefficients repeat this differencing operation. All these differences concern the presence of jobs, with their processing times held fixed.

M\"obius inversion recovers the allocation from these coefficients:
$$
x_{ij}^S(p_S)=\sum_{\substack{J\subseteq S\\j\in J}}G_{ij}^J(p_J)~.
$$
Our structural target is interaction order at most $m-1$, or equivalently,
\begin{equation*}
G_{ij}^J(p_J)=0\qquad\text{whenever }|J|\ge m~.
\end{equation*}
Under this condition, the expansion uses only sets of at most $m-1$ jobs. In particular, for two machines, $x_{ij}^S(p_S)=x_{ij}^{\{j\}}(p_{\{j\}})$: the allocation of each job depends only on its own column.

For a fixed instance $p$, the same expansion expresses each machine's load on $S$ as a multilinear polynomial of degree at most $m-1$ in the indicators of which jobs are present. The processing times remain fixed in this representation. This degree bound is the property used in the lower-bound construction.

\subsection{The High-Level Proof Framework}
\label{sec:two-step-framework}

As outlined in the technical overview, the proof of~\cref{thm:main} has two steps:
\begin{enumerate}[label=(\arabic*), leftmargin=*]
    \item \emph{Bounded-interaction reduction.}
    Replace any truthful fractional mechanism with a truthful fractional
    mechanism whose interaction order is at most $m-1$, while
    preserving its approximation guarantee up to an arbitrarily
    small additive loss.

    \item \emph{Hard-instance construction.}
    For any fractional allocation rule of interaction order at most
    $m-1$, construct a hard instance establishing the desired
    approximation lower bound.
\end{enumerate}

For clarity, we first give the hard-instance construction in Section~\ref{sec:bounded-interaction-lower-bound} and then establish the bounded-interaction reduction in Section~\ref{sec:interaction-elimination}. The construction shows how bounded interaction leads to an approximation lower bound, without using truthfulness. This explains why the reduction needs to produce a mechanism with bounded interaction. The reduction then uses truthfulness to obtain a mechanism to which the hard-instance construction applies. We state the corresponding lemmas in the same order.

\begin{restatable}[Lower bound under bounded interaction]{lemma}{boundedinteractionlowerbound}
\label{lem:bounded-interaction-lower-bound}
Fix $0\le\Gamma<(m+1)/2$. Every feasible fractional allocation family $x=(x^S)_S$ with interaction order at most $m-1$ admits an instance $(S,p)$ satisfying
$$
C\bigl(p,x^S(p)\bigr)>\Gamma\operatorname{OPT}(p)~.
$$
Moreover, there is a constant $L=L(m,\Gamma)\ge1$, independent of $x$, such that the instance can be chosen to satisfy
$$
\frac{|S|}{\operatorname{OPT}(p)}\le mL^2~.
$$
\end{restatable}


Truthfulness is used only in the \emph{bounded-interaction reduction} step. In the \emph{hard-instance construction} step, the bounded-interaction property alone yields the lower bound. The hard instances consist of replicated stars with an additional group of jobs cheap only at the center. Each edge job is cheap on the center and one leaf machine; all other processing times are large. An integral schedule assigning the extra group to the center and the edge groups to their respective leaves bounds the optimum. The bounded-degree representation links the center's load on singleton instances to its load on larger subsets. Combining these constraints yields the lower bound.

The next lemma uses truthfulness to reduce arbitrary mechanisms to the class covered by Lemma~\ref{lem:bounded-interaction-lower-bound}.

\begin{lemma}[Bounded interaction reduction]
\label{lem:bounded-interaction-replacement}
Let $M=(x^S,\pi^S)_S$ be a truthful, $\Gamma$-approximate fractional mechanism family. For every $\delta>0$, there exists a fractional mechanism family $M^\delta=(x^{S,\delta},\pi^{S,\delta})_S$ of interaction order at most $m-1$ such that
$$
C\bigl(p,x^{S,\delta}(p)\bigr)\le\Gamma\bigl(\operatorname{OPT}(p)+|S|\delta\bigr)
$$
for every instance $(S,p)$.
\end{lemma}

The construction first obtains a replacement invariant under adding jobs that have a zero-cost machine. This invariance supplies boundary identities for utility interactions. Truthfulness and complete allocation then yield differential identities, which eliminate higher-order interactions after compatible smoothing. The smoothed mechanism is not required to remain padding-invariant.

In the proof of Theorem~\ref{thm:main}, the bound on $|S|/\operatorname{OPT}(p)$ in Lemma~\ref{lem:bounded-interaction-lower-bound} lets us choose a fixed positive $\delta$ before constructing the replacement and its hard instance. No limit of the smoothed mechanisms is needed.

\begin{theorem}[Our main result]
\label{thm:main}
Every truthful, $\Gamma$-approximate fractional mechanism family satisfies
$$
\Gamma\ge\frac{m+1}{2}~.
$$
The same lower bound holds for truthful-in-expectation randomized mechanism families.
\end{theorem}

\begin{proof}
Let $M$ be a truthful, $\Gamma$-approximate fractional mechanism family. Suppose for a contradiction that $\Gamma<(m+1)/2$, and choose $\gamma$ with $\Gamma<\gamma<(m+1)/2$. Let $L=L(m,\gamma)$ be the constant from Lemma~\ref{lem:bounded-interaction-lower-bound}, and choose $\delta>0$ so that
$$
\Gamma mL^2\delta\le\gamma-\Gamma~.
$$
Apply Lemma~\ref{lem:bounded-interaction-replacement} at this fixed scale $\delta$. The resulting mechanism has interaction order at most $m-1$. Lemma~\ref{lem:bounded-interaction-lower-bound}, applied to this mechanism with parameter $\gamma$, gives an instance $(S,p)$ such that
$$
C\bigl(p,x^{S,\delta}(p)\bigr)>\gamma\operatorname{OPT}(p),
\qquad
\frac{|S|}{\operatorname{OPT}(p)}\le mL^2~.
$$
On this instance, however, the replacement guarantee gives
$$
\begin{aligned}
C\bigl(p,x^{S,\delta}(p)\bigr)
&\le\Gamma\bigl(\operatorname{OPT}(p)+|S|\delta\bigr)\\
&\le\bigl(\Gamma+\Gamma mL^2\delta\bigr)\operatorname{OPT}(p)
\le\gamma\operatorname{OPT}(p)~.
\end{aligned}
$$
This is a contradiction. Hence $\Gamma\ge(m+1)/2$.

The reduction in Section~\ref{sec:preliminaries} converts any TIE randomized mechanism family into a truthful fractional mechanism family with the same approximation guarantee. The fractional lower bound therefore proves the randomized claim.
\end{proof}

\section{Lower Bound under Bounded Interaction}
\label{sec:bounded-interaction-lower-bound}

We generate a random instance in three steps. First, we fix a \emph{star template}, consisting of a star graph together with a set of candidate jobs attached to each edge and to the center. In particular, the star has a center $v_0$ and $m-1$ leaves $v_1,\ldots,v_{m-1}$. For an integer $L\ge1$ to be chosen later, we associate a set $J_r$ of $L$ distinct jobs with each edge $\{v_0,v_r\}$, and a set $J_0$ of $L$ distinct jobs with the center. The sets $J_0,\ldots,J_{m-1}$ are pairwise disjoint, and we write $J=\bigcup_{r=0}^{m-1}J_r$ for the set of all $mL$ candidate jobs.

Intuitively, each vertex corresponds to a machine, each edge job is restricted to its two endpoint machines, and each job in $J_0$ is restricted to the center. In the second step, we randomize this correspondence by choosing a uniformly random bijection $\sigma:\{v_0,\ldots,v_{m-1}\}\to[m]$. For every edge job $j\in J_r$, $r\in[m-1]$, define its processing times by
$$
p_{ij}^{\sigma}=
\begin{cases}
1/L, & i\in\{\sigma(v_0),\sigma(v_r)\},\\
H/L, & \text{otherwise},
\end{cases}
$$
where $H>1$ is a sufficiently large finite parameter to be chosen later. For every job $j\in J_0$, set
$$
p_{ij}^{\sigma}=
\begin{cases}
1/L, & i=\sigma(v_0),\\
H/L, & \text{otherwise}.
\end{cases}
$$
These processing times implement the restricted-assignment picture with a large finite value $H/L$ in place of $\infty$. For each fixed $\sigma$, these definitions give a processing-time matrix $p^\sigma$ for all candidate jobs in $J$.

Finally, with the template and a sampled bijection $\sigma$ in place, we complete the construction using a fixed probability parameter $\rho\in[0,1]$. Independently of $\sigma$, we include each job in $J$, including every job in $J_0$, independently with probability $\rho$. Let $S_\rho\subseteq J$ be the resulting set of jobs. For each fixed $\rho$, this procedure defines a random instance $I_\rho=(S_\rho,p^\sigma_{S_\rho})$ with distribution $\mathcal D_\rho$, written $I_\rho\sim\mathcal D_\rho$. Its randomness comes from both the uniformly random bijection $\sigma$ and the independent selection of the job set $S_\rho$.

Fix $0\le\Gamma<(m+1)/2$ and a feasible fractional allocation family $x$ of interaction order at most $m-1$. For a realized instance $I=(S,p^\sigma_S)$, let $\operatorname{ALG}(I)$ denote the makespan achieved by $x$, and let $\operatorname{OPT}(I)$ denote the optimal integral makespan. Both values are zero on the empty instance. Our goal is to find a realized instance $I$ with $\operatorname{ALG}(I)>\Gamma\operatorname{OPT}(I)$.

We study $\mathbb E_{I_\rho\sim\mathcal D_\rho}[\operatorname{ALG}(I_\rho)]$ as a function of $\rho$; all randomness here comes from the input construction. To obtain the desired instance, it suffices to choose $L$ and $H$ and find a sampling probability $\rho>0$ such that
$$
\mathbb E_{I_\rho\sim\mathcal D_\rho}[\operatorname{ALG}(I_\rho)]
>\Gamma\mathbb E_{I_\rho\sim\mathcal D_\rho}[\operatorname{OPT}(I_\rho)]~.
$$
This inequality guarantees that some realization $I$ of $I_\rho$ satisfies $\operatorname{ALG}(I)>\Gamma\operatorname{OPT}(I)$. To establish the expected gap, we study the expected load of the center machine, which is a lower bound on the expected makespan.

\subsection{Overview of the Analysis}
\label{sec:star-analysis}

We analyze the expected load on the center machine $\sigma(v_0)$,
\begin{equation*}
F(\rho)
=\mathbb E_{\sigma,S_\rho}\left[
\frac1L\sum_{j\in S_\rho}
x_{\sigma(v_0),j}^{S_\rho}(p^\sigma_{S_\rho})
\right]~.
\end{equation*}
The expectation is over both the machine--vertex bijection and the independent job inclusions. We write $\operatorname{OPT}(\rho)=\mathbb E_{\sigma,S_\rho}[\operatorname{OPT}(I_\rho)]$. Since the center load is at most the makespan on every instance,
\begin{equation*}
F(\rho)\le\mathbb E[\operatorname{ALG}(I_\rho)]~.
\end{equation*}

Our analysis compares $F(\rho)$ and $\operatorname{OPT}(\rho)$ and explains the limitation imposed by bounded interaction. We first examine their behavior near the origin, where singleton instances, each containing exactly one job, determine the initial slopes. In the restricted-assignment picture, both slopes are $\Theta(m)$. As more jobs appear, however, OPT can assign jobs in $J_0$ to the center and each edge job to its corresponding leaf, processing all $m$ groups in parallel. Adding one job to each group increases this schedule's makespan by only $1/L$, even though $m$ jobs have been added. Once the job counts are sufficiently concentrated, this gives $\operatorname{OPT}(\rho)/\rho\le1+\varepsilon$. By increasing $L$, the ratio $\operatorname{OPT}(\rho)/\rho$ can thus fall from $m$ near the origin to at most $1+\varepsilon$ within an arbitrarily small neighborhood of the origin, and then stay at roughly the same constant scale as $\rho$ increases further. In this sense, OPT can roughly sustain this smaller average growth rate.

In the next part of the analysis, we show why bounded interaction prevents the same rapid decline in $F(\rho)/\rho$. The key observation is that $F(\rho)$ is a polynomial in $\rho$ whose degree is bounded solely by $m$. Since $F(0)=0$, the ratio $F(\rho)/\rho$ extends to a polynomial of bounded degree. Feasibility gives $0\le F(\rho)\le m\rho$, so this ratio takes values in $[0,m]$. Suppose we try to make $F(\rho)/\rho$ behave like $\operatorname{OPT}(\rho)/\rho$: drop by a fixed amount within a tiny neighborhood of the origin, and then settle at a much lower level. To make the initial drop so sharp, we need to adjust the polynomial's coefficients. But these same coefficients also control the curve over the rest of $[0,1]$; their effect does not stop when we leave that neighborhood. With the degree fixed, squeezing the drop into a sufficiently small neighborhood forces the curve to move too far somewhere else, either below $0$ or above $m$. Thus, we cannot make it drop arbitrarily quickly and then keep it within the allowed range. This restriction then yields $F(\rho)>\Gamma\operatorname{OPT}(\rho)$ for some $\rho>0$ when $\Gamma<(m+1)/2$ and $L$ and $H$ are sufficiently large.

\subsection{Behavior near the Origin: Comparable Initial Growth}
Fix $L$ and $H$, and let $\rho\downarrow0$. The instance is empty with probability tending to one. Conditioned on being nonempty, however, it contains exactly one job with probability tending to one.

First consider a singleton edge job $j\in J_r$, where $r\in[m-1]$. In the restricted-assignment picture, it can use only its two endpoint machines. Because we assign machine indices to the star vertices uniformly at random, either endpoint machine is equally likely to be the center. Swapping their roles leaves the job and its processing-time vector unchanged, so it also leaves the algorithm's allocation unchanged. Thus, regardless of how the algorithm divides the job between the two endpoints, its expected center load is $1/(2L)$.

Now we consider a singleton job $j\in J_0$. It must be assigned entirely to the center, contributing $1/L$ to its load. The $L$ jobs in $J_0$ therefore contribute $1$ to the initial slope, in addition to the $(m-1)/2$ contributed by the $(m-1)L$ edge jobs. Together, these contributions give $F'(0)=(m+1)/2$ in this picture. For comparison, all $mL$ singleton instances have integral optimum $1/L$, giving $\operatorname{OPT}'(0)=m$. Thus, the optimal solution has no advantage in this regime. The following lemma formalizes this argument for finite $H$.

\begin{lemma}
\label{lem:star-origin}
Either a singleton instance $I$ satisfies $\operatorname{ALG}(I)>\Gamma\operatorname{OPT}(I)$, or the expected center load satisfies
$$
F'(0)\ge\frac{m+1}{2}-\frac{m(m-1)\Gamma}{2H}~.
$$
\end{lemma}

\begin{proof}
If a singleton instance satisfies $\operatorname{ALG}(I)>\Gamma\operatorname{OPT}(I)$, the first conclusion holds. Otherwise, every singleton instance has makespan at most $\Gamma/L$, since its integral optimum is $1/L$. We prove the stated slope bound in this case.

For a fixed bijection $\sigma$ and job set $S\subseteq J$, write
$$
f_\sigma(S)=\frac1L\sum_{j\in S}x_{\sigma(v_0),j}^S(p^\sigma_S)~.
$$
We have $f_\sigma(\varnothing)=0$ and $0\le f_\sigma(S)\le |S|/L$. Each singleton $\{j\}$ occurs with probability $\rho(1-\rho)^{|J|-1}$, so
$$
F(\rho)=\rho(1-\rho)^{|J|-1}
\sum_{j\in J}\mathbb E_\sigma[f_\sigma(\{j\})]+R_F(\rho)~,
$$
where $R_F(\rho)$ is the contribution from instances with at least two jobs. Since $|S_\rho|\le |S_\rho|(|S_\rho|-1)$ whenever $|S_\rho|\ge2$, we have
$$
0\le R_F(\rho)
\le\frac1L\mathbb E\bigl[|S_\rho|(|S_\rho|-1)\bigr]
=\frac{|J|(|J|-1)}{L}\rho^2~.
$$
The last equality follows by summing over ordered pairs of distinct jobs, each of which appears with probability $\rho^2$. Dividing the expression for $F(\rho)$ by $\rho$ and taking $\rho\downarrow0$ gives
$$
F'(0)=\sum_{j\in J}\mathbb E_\sigma[f_\sigma(\{j\})]~.
$$

We now bound these singleton contributions. First fix an edge job $j\in J_r$, $r\in[m-1]$, and let $y_i^\sigma=x_{ij}^{\{j\}}(p^\sigma_{\{j\}})$. Write $\beta_j^\sigma$ for the total fraction assigned to machines other than $\sigma(v_0)$ and $\sigma(v_r)$. The singleton makespan bound gives $(H/L)y_i^\sigma\le\Gamma/L$ for each such machine. Hence
$$
0\le\beta_j^\sigma\le\frac{(m-2)\Gamma}{H}~.
$$
Pair $\sigma$ with the bijection obtained by swapping the machines at $v_0$ and $v_r$. This swap leaves the singleton input unchanged: in particular, none of the jobs in $J_0$ is present. Thus the allocation vector $y^\sigma$ is unchanged, while the center becomes $\sigma(v_r)$. Since $\sigma$ is uniform over all bijections,
$$
2\mathbb E_\sigma[y_{\sigma(v_0)}^\sigma]
=\mathbb E_\sigma[y_{\sigma(v_0)}^\sigma+y_{\sigma(v_r)}^\sigma]
=1-\mathbb E_\sigma[\beta_j^\sigma]~.
$$
Thus $\mathbb E_\sigma[f_\sigma(\{j\})]\ge[1-(m-2)\Gamma/H]/(2L)$. For a job $j\in J_0$, the same singleton guarantee gives $x_{ij}^{\{j\}}(p^\sigma_{\{j\}})\le\Gamma/H$ on each of the $m-1$ machines other than the center. Feasibility therefore gives
$$
\mathbb E_\sigma[f_\sigma(\{j\})]
\ge\frac1L\left(1-\frac{(m-1)\Gamma}{H}\right)~.
$$
Summing over the $L$ jobs in $J_0$ and the $(m-1)L$ edge jobs yields
$$
\begin{aligned}
F'(0)
&\ge1-\frac{(m-1)\Gamma}{H}
+\frac{m-1}{2}\left(1-\frac{(m-2)\Gamma}{H}\right)\\
&=\frac{m+1}{2}-\frac{m(m-1)\Gamma}{2H}~,
\end{aligned}
$$
as claimed.
\end{proof}

\subsection{Moving beyond the Origin: The Limitation of Bounded Interaction}
\label{sec:bounded-interaction-limitation}
As $\rho$ increases, OPT can exploit parallel processing by assigning every included job in $J_0$ to the center and every included edge job in $J_r$, $r\in[m-1]$, to the corresponding leaf $\sigma(v_r)$. As a result, adding one job to each of the $m$ groups increases this schedule's makespan by only $1/L$. In our random instance, each machine receives $\rho L$ jobs in expectation and hence has expected load $\rho$. When $\rho L$ is sufficiently large, the job counts are concentrated around their means, so this schedule's expected makespan is close to $\rho$. The new group $J_0$ uses the center while the edge groups use the leaves, spreading the expected total work $m\rho$ over all $m$ machines. The following lemma bounds the imbalance between these groups and gives the upper bound on $\operatorname{OPT}(\rho)$ that we need.

\begin{lemma}
\label{lem:star-optimum}
For every $\rho\in[0,1]$,
$
\operatorname{OPT}(\rho)
\le\rho+m\sqrt{\frac{\rho(1-\rho)}{L}}~.
$
Consequently, for every $\varepsilon>0$ and every $\rho\in(0,1]$ satisfying $\rho\ge m^2/(\varepsilon^2L)$, we have
$$
\operatorname{OPT}(\rho)\le(1+\varepsilon)\rho~.
$$
\end{lemma}

\begin{proof}
For each $r\in\{0,\ldots,m-1\}$, let $X_r=|S_\rho\cap J_r|$. Assigning the jobs in $J_0$ to the center and the jobs in each $J_r$, $r\in[m-1]$, to their leaf gives a feasible integral schedule, so $\operatorname{OPT}(I_\rho)\le\max_{0\le r\le m-1}X_r/L$ for every realization of $\sigma$ and $S_\rho$. Each $X_r$ has the binomial distribution with parameters $L$ and $\rho$. Thus $\mathbb E[X_r/L]=\rho$ and $\operatorname{Var}(X_r/L)=\rho(1-\rho)/L$. It follows that
$$
\begin{aligned}
\operatorname{OPT}(\rho)
&\le\mathbb E\left[\max_{0\le r\le m-1}\frac{X_r}{L}\right]\\
&\le\rho+\sum_{r=0}^{m-1}\mathbb E\left[\left|\frac{X_r}{L}-\rho\right|\right]\\
&\le\rho+m\sqrt{\frac{\rho(1-\rho)}{L}}~,
\end{aligned}
$$
where the last inequality follows from Cauchy--Schwarz. For $\rho>0$, dividing by $\rho$ and using $1-\rho\le1$ gives $\operatorname{OPT}(\rho)/\rho\le1+m/\sqrt{L\rho}$. Under the stated condition on $\rho$, the second term is at most $\varepsilon$, proving the consequence.
\end{proof}

Lemma~\ref{lem:star-optimum} shows that $\operatorname{OPT}(\rho)/\rho$ can fall to a constant scale within a neighborhood of the origin that shrinks as $L$ grows. We now show why bounded interaction prevents $F(\rho)/\rho$ from decreasing by a fixed amount on an arbitrarily small scale. The first step is to express $F$ as a polynomial whose degree depends only on $m$.

\begin{lemma}[Polynomial representation]
\label{lem:star-polynomial}
If $x$ has interaction order at most $m-1$, then $F$ is a polynomial of degree at most $m-1$. Moreover, $F(0)=0$ and
$$
0\le F(\rho)\le m\rho
\qquad\text{for every }\rho\in[0,1]~.
$$
\end{lemma}

\begin{proof}
Fix the machine--vertex bijection $\sigma$. By M\"obius inversion and the absence of interactions of order at least $m$, the center load on a job set $S\subseteq J$ is
$$
\frac1L\sum_{j\in S}x_{\sigma(v_0),j}^{S}(p_S^\sigma)
=\frac1L\sum_{\substack{\varnothing\ne T\subseteq S\\|T|\le m-1}}
\sum_{j\in T}G_{\sigma(v_0),j}^{T}(p_T^\sigma)~.
$$
Each fixed set $T\subseteq J$ is contained in $S_\rho$ with probability $\rho^{|T|}$, since the jobs appear independently. Averaging over $S_\rho$ and then over $\sigma$ gives
$$
F(\rho)
=\frac1L\sum_{\substack{\varnothing\ne T\subseteq J\\|T|\le m-1}}
\rho^{|T|}\,
\mathbb E_\sigma\left[\sum_{j\in T}G_{\sigma(v_0),j}^{T}(p_T^\sigma)\right]~.
$$
The coefficients in this expression do not depend on $\rho$. Hence $F$ has degree at most $m-1$ and has no constant term. Finally, feasibility gives $0\le x_{\sigma(v_0),j}^{S}\le1$, so
$$
0\le F(\rho)\le\frac1L\,\mathbb E[|S_\rho|]
=\frac{|J|}{L}\rho=m\rho~.
$$
This proves the remaining bounds.
\end{proof}

The degree bound and the feasibility bounds work together. Since $F(0)=0$, the ratio $F(\rho)/\rho$ extends to a polynomial of degree at most $m-2$, with value $F'(0)$ at the origin and values in $[0,m]$ throughout $[0,1]$. We use this continuous extension whenever we refer to the ratio at the origin. A polynomial of fixed degree is determined by its values at finitely many fixed points. Keeping those values bounded also limits how quickly the polynomial can decrease near the origin.

For a simple example, suppose that $F(\rho)/\rho$ is linear and fix its initial value $a=F'(0)>0$. Its remaining freedom is the endpoint value $F(1)$: we have $F(\rho)/\rho=a+(F(1)-a)\rho$. Since $F(1)\ge0$, the fastest possible decrease is attained by $F(\rho)/\rho=a(1-\rho)$. Any steeper line would become negative before $\rho=1$. Thus, the ratio cannot lose a fixed fraction of its initial value within an arbitrarily short interval, whereas increasing $L$ allows $\operatorname{OPT}(\rho)/\rho$ to decrease on an arbitrarily small scale.

Now suppose that $F(\rho)/\rho$ is quadratic. Once its value $F'(0)$ at the origin is fixed, its values at $\rho=1/2$ and $\rho=1$ determine the entire curve. In particular,
$$
\left.\frac{d}{d\rho}\left(\frac{F(\rho)}{\rho}\right)\right|_{\rho=0}
=8F(1/2)-3F'(0)-F(1)~.
$$
This identity shows that the initial slope cannot be chosen independently of the curve's values farther from the origin. With $F'(0)$ fixed, making the slope more negative requires decreasing the ratio at $\rho=1/2$ or increasing it at $\rho=1$. However, these values cannot be adjusted arbitrarily: the entire curve must remain in $[0,m]$. Trying to force a sharper initial decline eventually pushes some part of the curve below zero or above $m$.

\begin{figure}[t]
\centering
\includegraphics[width=.86\linewidth]{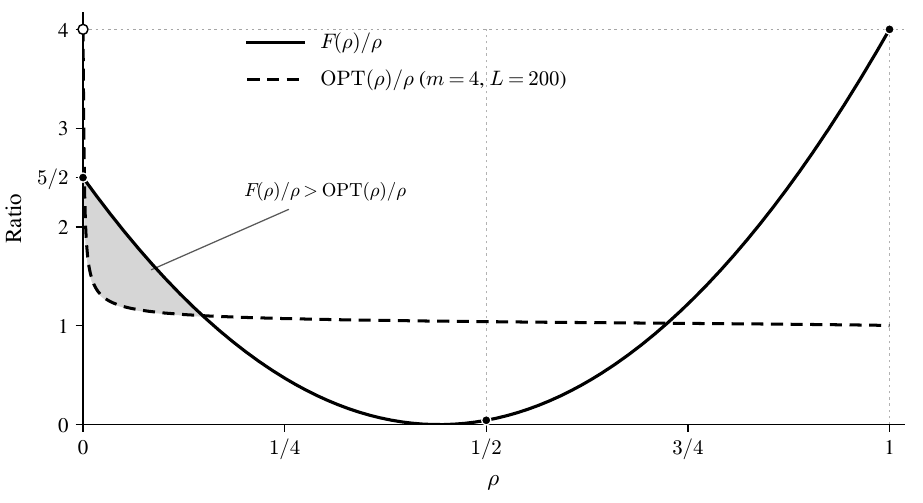}
\caption{The solid curve is the quadratic candidate $F(\rho)/\rho=\frac52[1-(1+\sqrt{8/5})\rho]^2$. The dashed curve is the expected integral optimum ratio $\operatorname{OPT}(\rho)/\rho$ for the star construction including $J_0$, with $m=4$, $L=200$, and $H=1000$, computed from the finite binomial distribution. Filled dots mark the solid curve at $\rho=0,1/2,1$. The shaded region indicates $F(\rho)/\rho>\operatorname{OPT}(\rho)/\rho$. Values at the origin are interpreted by continuity.}
\label{fig:quadratic-center-ratio}
\end{figure}

Figure~\ref{fig:quadratic-center-ratio} makes this restriction concrete. We take $m=4$ and fix the illustrative initial value $F'(0)=5/2$. Among quadratic ratios that stay in $[0,4]$ on $[0,1]$, the solid curve has the fastest possible initial decline. Nevertheless, the dashed curve $\operatorname{OPT}(\rho)/\rho$ decreases more quickly and soon falls below it. In the shaded region, we have $F(\rho)/\rho>\operatorname{OPT}(\rho)/\rho$, or equivalently $F(\rho)>\operatorname{OPT}(\rho)$. Since the algorithm's expected makespan is at least $F(\rho)$, this is the type of gap we seek.

The key distinction is that the restriction on the polynomial depends only on its degree and allowed range, whereas increasing $L$ can move OPT's decline arbitrarily close to the origin. Thus, OPT can reach a much smaller ratio while the polynomial is still close to its initial value. We now give a formal proof of this observation.

\begin{lemma}[Limited decrease near the origin]
\label{lem:star-limited-decrease}
There is a constant $C_m\ge0$, depending only on $m$, such that for every choice of $L$ and $H$ and every feasible allocation family of interaction order at most $m-1$, the expected center load satisfies
$$
F(\rho)\ge\rho F'(0)-C_m\rho^2
\qquad\text{for every }\rho\in[0,1]~.
$$
\end{lemma}

\begin{proof}
The degree bound and the identity $F(0)=0$ in Lemma~\ref{lem:star-polynomial} imply that the ratio $F(\rho)/\rho$ extends to a polynomial of degree at most $d=m-2$, with value $F'(0)$ at the origin. If $d=0$, then the ratio is constant, so $F(\rho)=\rho F'(0)$ and the claim holds with $C_m=0$.

Suppose that $d\ge1$. Interpolate $F(\rho)/\rho$ at the fixed points $0,1/d,\ldots,1$, using its continuous extension at the origin. The Lagrange interpolation formula gives
$$
\frac{F(\rho)}{\rho}
=F'(0)\ell_0(\rho)+\sum_{r=1}^{d}\frac{F(r/d)}{r/d}\ell_r(\rho)
\qquad\text{for }\rho>0~,
$$
where
$$
\ell_r(\rho)=\prod_{\substack{0\le s\le d\\s\ne r}}
\frac{\rho-s/d}{r/d-s/d}~.
$$
The polynomials $\ell_0,\ldots,\ell_d$ depend only on $d$. In particular,
$$
C_m:=m\sum_{r=0}^{d}\max_{t\in[0,1]}|\ell_r'(t)|
$$
is finite and depends only on $m$, not on $L$, $H$, or the allocation family. We now use the feasibility bound $0\le F(\rho)\le m\rho$ from Lemma~\ref{lem:star-polynomial}. It bounds every sampled ratio between $0$ and $m$, including $F'(0)$ at the origin by continuity. Differentiating the interpolation formula therefore yields
$$
\begin{aligned}
\left|\frac{d}{d\rho}\left(\frac{F(\rho)}{\rho}\right)\right|
&\le |F'(0)|\,|\ell_0'(\rho)|
+\sum_{r=1}^{d}\left|\frac{F(r/d)}{r/d}\right|\,|\ell_r'(\rho)|\\
&\le C_m
\qquad\text{for every }\rho\in(0,1]~.
\end{aligned}
$$
Integrating from the origin gives $F(\rho)/\rho\ge F'(0)-C_m\rho$ for $\rho>0$. Multiplying by $\rho$ proves the claim for positive $\rho$; at $\rho=0$, it follows from $F(0)=0$.
\end{proof}

For $\rho>0$, Lemma~\ref{lem:star-limited-decrease} gives $F(\rho)/\rho\ge F'(0)-C_m\rho$. Thus, for any fixed $\varepsilon>0$, the ratio is at least $F'(0)-\varepsilon$ whenever $0<\rho\le\min\{1,\varepsilon/(1+C_m)\}$. This neighborhood depends only on $m$ and $\varepsilon$; it does not shrink as $L$ grows. Meanwhile, Lemma~\ref{lem:star-optimum} puts $\operatorname{OPT}(\rho)/\rho$ below $1+\varepsilon$ once $\rho\ge m^2/(\varepsilon^2L)$. Increasing $L$ therefore allows the optimum's ratio to reach this smaller scale while $F(\rho)/\rho$ is still at least $F'(0)-\varepsilon$.

We now combine these bounds to prove Lemma~\ref{lem:bounded-interaction-lower-bound}, restated below.

\boundedinteractionlowerbound*

\begin{proof}
We first choose the parameters of the random-instance construction independently of the allocation family and establish the claimed ratio bound for every nonempty instance in its support. We then fix an arbitrary allocation family and show that at least one of these instances violates its $\Gamma$-approximation guarantee. The same instance will therefore satisfy both conclusions.

It suffices to consider $1\le\Gamma<(m+1)/2$, since the case $\Gamma<1$ follows from the result with $\Gamma=1$. Choose $\varepsilon>0$ sufficiently small that
$$
\frac{m+1}{2}-2\varepsilon>\Gamma(1+\varepsilon)~.
$$
Choose $H>1$ such that $m(m-1)\Gamma/(2H)\le\varepsilon$, and fix
$$
\rho=\min\left\{\frac12,\frac{\varepsilon}{1+C_m}\right\}>0~.
$$
Next, choose an integer $L\ge1$ with $L\ge m^2/(\varepsilon^2\rho)$. All parameters depend only on $m$ and $\Gamma$. Every nonempty instance $(S,p)$ generated from this template has $|S|\le mL$ and $\operatorname{OPT}(p)\ge1/L$, since every processing time is at least $1/L$. Consequently, every such instance satisfies
$$
\frac{|S|}{\operatorname{OPT}(p)}\le mL^2~.
$$

It remains to find a violating instance within this support. Fix an allocation family $x$ as in the lemma. If a singleton instance satisfies $C(p,x^S(p))>\Gamma\operatorname{OPT}(p)$, we are done. Otherwise, Lemma~\ref{lem:star-origin} and the choice of $H$ give
$$
F'(0)\ge\frac{m+1}{2}-\frac{m(m-1)\Gamma}{2H}
\ge\frac{m+1}{2}-\varepsilon~.
$$
Lemmas~\ref{lem:star-limited-decrease} and~\ref{lem:star-optimum} give
$$
\begin{aligned}
\frac{F(\rho)}{\rho}
&\ge F'(0)-C_m\rho
\ge\frac{m+1}{2}-2\varepsilon\\
&>\Gamma(1+\varepsilon)
\ge\Gamma\frac{\operatorname{OPT}(\rho)}{\rho}~.
\end{aligned}
$$
Multiplying by $\rho>0$ and using that the expected center load is at most the expected makespan gives
$$
\mathbb E[\operatorname{ALG}(I_\rho)]
\ge F(\rho)>\Gamma\mathbb E[\operatorname{OPT}(I_\rho)]~.
$$
Hence some realized instance $I$ satisfies $\operatorname{ALG}(I)>\Gamma\operatorname{OPT}(I)$. This instance is nonempty, since both makespans are zero on the empty instance, and thus also satisfies the ratio bound established above. This proves both conclusions.
\end{proof}

\section{Eliminating Higher-Order Interactions}
\label{sec:interaction-elimination}

In this section, we prove Lemma~\ref{lem:bounded-interaction-replacement}. Starting from a truthful fractional mechanism family with approximation guarantee $\Gamma$, for every fixed $\delta>0$ we construct a fractional mechanism family of interaction order at most $m-1$ whose makespan is at most $\Gamma(\operatorname{OPT}(p)+|S|\delta)$ for every finite job set $S$ and every nonnegative report $p$ on $S$.

The additive loss $\Gamma|S|\delta$ comes from a smoothing operation. We first present the argument under some smoothness conditions, requiring only continuity of the allocation rules, and then remove this assumption by smoothing at the end of the section.

\subsection{Truthfulness and Utility Interactions}
\label{sec:utility-interactions}

The standard subgradient characterization of truthfulness (see Archer and Kleinberg~\cite{DBLP:journals/geb/ArcherK14}) states that, in our cost setting, $-x_i^S(p)$ is a subgradient of $u_i^S(\cdot,p_{-i})$ at $p_i$. Continuity of the allocation rule then gives the derivative identity below. We include a proof for completeness.

\begin{lemma}[Utility derivatives]
\label{lem:utility-derivatives}
Let $x$ and $u$ be the allocation rules and normalized utilities of a truthful mechanism family, and suppose that each $x^S$ is continuous on strictly positive reports. For every finite job set $S$, machine $i$, and fixed positive $p_{-i}$, the function $u_i^S(\cdot,p_{-i})$ is continuously differentiable on $\mathbb R_{>0}^S$, with
\begin{equation*}
\frac{\partial u_i^S(p)}{\partial p_{ij}}
=-x_{ij}^S(p)
\qquad(j\in S)~.
\end{equation*}
\end{lemma}

\begin{proof}
Fix a finite job set $S$ and a machine $i$. For any two nonnegative reports $p=(p_i,p_{-i})$ and $q=(q_i,p_{-i})$ that differ only in machine $i$'s row, truthfulness and the definition of normalized utility give
\begin{equation}
u_i^S(p_i,p_{-i})
\ge
u_i^S(q_i,p_{-i})
+(q_i-p_i)\cdot x_i^S(q_i,p_{-i})~.
\label{eq:normalized-ic}
\end{equation}
This incentive inequality holds without any continuity assumption.

Now fix $j\in S$ and increase only the processing time $p_{ij}$ by $h>0$, keeping all other entries unchanged. Write the resulting report as $p+he_{ij}$. Applying \eqref{eq:normalized-ic} in both directions gives
\begin{equation*}
-hx_{ij}^S(p)
\le u_i^S(p+he_{ij})-u_i^S(p)
\le -hx_{ij}^S(p+he_{ij})~.
\end{equation*}
Similarly, at a strictly positive report $p$, decreasing only $p_{ij}$ by $0<h<p_{ij}$ gives
\begin{equation*}
-hx_{ij}^S(p-he_{ij})
\le u_i^S(p)-u_i^S(p-he_{ij})
\le -hx_{ij}^S(p)~.
\end{equation*}
Assuming that $x^S$ is continuous on strictly positive reports, dividing both inequalities by $h$ and letting $h\downarrow0$ gives
\begin{equation*}
\left.
\begin{aligned}
\lim_{h\downarrow0}
\frac{u_i^S(p+he_{ij})-u_i^S(p)}{h}
&=-x_{ij}^S(p),\\
\lim_{h\downarrow0}
\frac{u_i^S(p)-u_i^S(p-he_{ij})}{h}
&=-x_{ij}^S(p)
\end{aligned}
\right\}
\quad\Longrightarrow\quad
\frac{\partial u_i^S(p)}{\partial p_{ij}}
=-x_{ij}^S(p)~.
\end{equation*}
For every fixed positive $p_{-i}$, these partial derivatives are continuous in $p_i$, so $u_i^S(\cdot,p_{-i})$ is continuously differentiable on $\mathbb R_{>0}^S$.
\end{proof}

Thus the allocation of job $j$ is the negative slope of utility in that job's processing time. We next record utility bounds that hold without any continuity assumption. For a report $p$ on a finite job set $S$, write $p^{i,0}=(0_i,p_{-i})$ for the report obtained by setting machine $i$'s entire row to zero. Recall that the normalized utility satisfies $u_i^S(p^{i,0})=0$.

\begin{lemma}[Utility bounds]
\label{lem:utility-bounds}
Let $x$ and $u$ be the allocation rules and normalized utilities of a truthful mechanism family. For every finite job set $S$, machine $i$, and nonnegative report $p$,
\begin{equation*}
-\sum_{j\in S}p_{ij}
\le u_i^S(p)
\le -p_i\cdot x_i^S(p)
\le 0~.
\end{equation*}
\end{lemma}

\begin{proof}
Fix a finite job set $S$, a machine $i$, and a nonnegative report $p$. Taking $q_i=0$ in \eqref{eq:normalized-ic} and using the zero-row normalization yields
\begin{equation*}
u_i^S(p)
\ge -p_i\cdot x_i^S(p^{i,0})
\ge -\sum_{j\in S}p_{ij}~.
\end{equation*}
Conversely, applying \eqref{eq:normalized-ic} with true row $0$ and reported row $p_i$ gives
\begin{equation*}
0=u_i^S(p^{i,0})
\ge u_i^S(p)+p_i\cdot x_i^S(p)
\quad\Longrightarrow\quad
u_i^S(p)\le -p_i\cdot x_i^S(p)\le 0~.
\end{equation*}
These inequalities prove the claimed utility bounds.
\end{proof}

Any feasible family $(x,u)$ satisfying \eqref{eq:normalized-ic} and the zero-row normalization is truthfully implemented by the finite payments $\pi_i^S(p)=u_i^S(p)+p_i\cdot x_i^S(p)$. We will use this observation when constructing replacements.

We now define the utility counterpart of the allocation interaction from Definition~\ref{def:job-interaction}. For a finite job set $J$, let
\begin{equation*}
V_i^J(p_J)=\sum_{S\subseteq J}(-1)^{|J|-|S|}u_i^S(p_S)~.
\end{equation*}
For example, $V_i^{\{a,b\}}=u_i^{\{a,b\}}-u_i^{\{a\}}-u_i^{\{b\}}$. It measures the part of utility that remains after subtracting the contributions of smaller job sets.

Under the temporary continuity assumption on $x$, Lemma~\ref{lem:utility-derivatives} gives $\partial u_i^S/\partial p_{ij}=-x_{ij}^S$ at every strictly positive report. Differentiating the defining sum for $V_i^J$ then gives
$$
\frac{\partial V_i^J}{\partial p_{ij}}
=-\sum_{\substack{S\subseteq J\\j\in S}}(-1)^{|J|-|S|}x_{ij}^S
=-G_{ij}^J~.
$$
The terms with $j\notin S$ have derivative zero.

Recall that the goal of this section is to construct a replacement satisfying $G_{ij}^J=0$ for every $J$ with $|J|\ge m$, every machine $i$, and every job $j\in J$. Under the continuity assumption on $x$, the identity above reduces this goal on strictly positive reports to proving $\partial V_i^J/\partial p_{ij}=0$. In fact, for a suitable replacement, we will prove the stronger statement that $V_i^J(p_J)=0$ for every $i$, every $J$ with $|J|\ge m$, and every strictly positive report $p_J$. For each such $J$, the argument uses two main conditions:
\begin{enumerate}[label=(\arabic*)]
\item \emph{Zero boundary condition.} For every machine $i$, $V_i^J(p_J)=0$ whenever $p_{ik}=0$ for some $k\in J$.
\item \emph{Differential conservation law.} At every strictly positive report, for each $j\in J$,
$$
\sum_{i=1}^m\frac{\partial V_i^J(p_J)}{\partial p_{ij}}=0~.
$$
\end{enumerate}
Together with the bounds inherited from utilities, these conditions will force $V_i^J$ to vanish (i.e., $V_i^J(p_J)=0$ for every strictly positive report). The conservation law follows directly from complete allocation and the derivative relation above.

\begin{lemma}[Differential conservation law]
\label{lem:utility-interaction-conservation}
Assume that the allocation rules $x^S$ of a truthful mechanism family are continuous on strictly positive reports. Then, for every $J$ with $|J|\ge2$, every strictly positive report $p_J$, and every $j\in J$,
$$
\sum_{i=1}^m\frac{\partial V_i^J(p_J)}{\partial p_{ij}}=0~.
$$
\end{lemma}

\begin{proof}
Complete allocation gives
\begin{equation}
\begin{aligned}
\sum_{i=1}^m G_{ij}^J(p_J)
&=\sum_{i=1}^m\sum_{\substack{S\subseteq J\\j\in S}}
(-1)^{|J|-|S|}x_{ij}^S(p_S)\\
&=\sum_{\substack{S\subseteq J\\j\in S}}
(-1)^{|J|-|S|}\underbrace{\sum_{i=1}^m x_{ij}^S(p_S)}_{=1}\\
&=\sum_{\substack{S\subseteq J\\j\in S}}(-1)^{|J|-|S|}
=(1-1)^{|J|-1}=0~.
\end{aligned}
\label{eq:interaction-conservation}
\end{equation}
This identity holds without assuming continuity of the allocation rules. Under the stated continuity assumption, substituting $\partial V_i^J/\partial p_{ij}=-G_{ij}^J$ proves the claim.
\end{proof}

The boundary condition need not hold for an arbitrary truthful mechanism. In Section~\ref{sec:padding-replacement}, we construct a truthful \emph{padding-invariant replacement} with the same approximation guarantee. A padding job has zero processing time on at least one machine. For the replacement, adding or removing such a job leaves the allocations of all other jobs unchanged. If the padding job $k$ has $p_{ik}=0$, machine $i$'s normalized utility is also unchanged, which yields the boundary condition. Section~\ref{sec:interaction-boundary-smoothing} then combines the two conditions to prove $V_i^J=0$ under continuity of the allocation rules, and we remove that assumption by smoothing afterward.

\subsection{A Padding-Invariant Replacement}
\label{sec:padding-replacement}

In this subsection, we introduce a \emph{padding-appending operation} that transforms an original mechanism family into a new one. Fix a truthful fractional mechanism family with allocation rules $x$ and normalized utilities $u$. A \emph{padding job} is specified by a processing-time vector $v\in\mathbb R_{\ge0}^m$ satisfying $\min_i v_i=0$; that is, it has zero processing time on at least one machine. Fix such a vector $v$ independently of the input reports.

Given an input instance with job set $S$ and processing-time matrix $p$, choose a fresh job label $k\notin S$ independently of $p$. We construct an enlarged instance with job set $S\cup\{k\}$ by adding a job $k$ whose processing time on machine $i$ is $v_i$. Write $p\oplus v$ for its processing-time matrix: it agrees with $p$ on all jobs in $S$ and has the additional column $(p\oplus v)_{ik}=v_i$ for every $i\in[m]$.

We then run the original mechanism on this enlarged instance and keep only its allocation of the jobs in $S$. Thus the output on the original input assigns each job $j\in S$ exactly as the original mechanism would assign it in the presence of the additional job $k$. We call the resulting mechanism the \emph{padding-appended mechanism} associated with the original mechanism and $v$. Denote the operation by $T_v$; its allocation rules and normalized utilities are defined by
\begin{equation*}
\begin{aligned}
[T_v(x)]_{ij}^S(p)&=x_{ij}^{S\cup\{k\}}(p\oplus v),\\
[T_v(u)]_i^S(p)&=u_i^{S\cup\{k\}}(p\oplus v)
                 -u_i^{S\cup\{k\}}(p^{i,0}\oplus v)~.
\end{aligned}
\end{equation*}
In the second line, $p^{i,0}\oplus v$ sets machine $i$'s processing times for the original jobs in $S$ to zero while keeping the appended cost $v_i$ fixed. This reference utility is independent of machine $i$'s report on $S$, and subtracting it makes the transformed utility zero at a zero own row. The payments are recovered from these allocations and normalized utilities as described in Section~\ref{sec:utility-interactions}.

\begin{lemma}
\label{lem:padding-preservation}
Let $(x,u)$ be the allocation rules and normalized utilities of a truthful, $\Gamma$-approximate fractional mechanism family. For every fixed padding job $v\in\mathbb R_{\ge0}^m$ with $\min_i v_i=0$, the transformed family $(T_v(x),T_v(u))$ satisfies:
\begin{enumerate}[label=(\arabic*)]
\item \emph{Truthfulness.} For every finite job set $S$, machine $i$, and two nonnegative reports $p=(p_i,p_{-i})$ and $q=(q_i,p_{-i})$,
\begin{equation*}
[T_v(u)]_i^S(p)
\ge [T_v(u)]_i^S(q)
+(q_i-p_i)\cdot[T_v(x)]_i^S(q)~.
\end{equation*}
Together with zero-row normalization, these inequalities implement $T_v(x)$ truthfully with finite payments.
\item \emph{Approximation guarantee.} For every finite nonempty job set $S$ and every nonnegative report $p$ with $\operatorname{OPT}(p)>0$,
\begin{equation*}
C\bigl(p,[T_v(x)]^S(p)\bigr)
\le \Gamma\operatorname{OPT}(p)~.
\end{equation*}
\end{enumerate}
\end{lemma}

\begin{proof}
Fix a padding job $v$. For each finite job set $S$, choose a fresh label $k\notin S$ independently of the reports. Restricting the original allocation on $S\cup\{k\}$ to $S$ preserves complete allocation of every job in $S$, so $T_v(x)$ is feasible.

For truthfulness, fix a machine $i$ and two nonnegative reports $p=(p_i,p_{-i})$ and $q=(q_i,p_{-i})$ on $S$. The enlarged reports $p\oplus v$ and $q\oplus v$ have the same opponents' rows, while machine $i$'s rows are $(p_i,v_i)$ and $(q_i,v_i)$, respectively. Applying the normalized incentive inequality \eqref{eq:normalized-ic} of the original mechanism on $S\cup\{k\}$ gives
\begin{equation*}
\begin{aligned}
u_i^{S\cup\{k\}}(p\oplus v)
&\ge u_i^{S\cup\{k\}}(q\oplus v)
+\sum_{j\in S}(q_{ij}-p_{ij})x_{ij}^{S\cup\{k\}}(q\oplus v)\\
&\qquad +(v_i-v_i)x_{ik}^{S\cup\{k\}}(q\oplus v)\\
&=u_i^{S\cup\{k\}}(q\oplus v)
+(q_i-p_i)\cdot[T_v(x)]_i^S(q)~.
\end{aligned}
\end{equation*}
The appended coordinate contributes zero because its processing time is fixed at $v_i$. Moreover, $p^{i,0}=q^{i,0}$, so the reference utilities subtracted in the definitions of $[T_v(u)]_i^S(p)$ and $[T_v(u)]_i^S(q)$ are equal. Subtracting this common reference yields
\begin{equation*}
\begin{aligned}
[T_v(u)]_i^S(p)
&=u_i^{S\cup\{k\}}(p\oplus v)
-u_i^{S\cup\{k\}}(p^{i,0}\oplus v)\\
&\ge u_i^{S\cup\{k\}}(q\oplus v)
-u_i^{S\cup\{k\}}(q^{i,0}\oplus v)
+(q_i-p_i)\cdot[T_v(x)]_i^S(q)\\
&=[T_v(u)]_i^S(q)
+(q_i-p_i)\cdot[T_v(x)]_i^S(q)~.
\end{aligned}
\end{equation*}
The transformed utilities are finite and satisfy $[T_v(u)]_i^S(p^{i,0})=0$. Hence the finite payments
\begin{equation*}
\pi_i^{S,v}(p)
=[T_v(u)]_i^S(p)+p_i\cdot[T_v(x)]_i^S(p)
\end{equation*}
implement $T_v(x)$ truthfully.

For the approximation guarantee, fix a nonempty job set $S$ and a nonnegative report $p$ with $\operatorname{OPT}(p)>0$. An optimal integral schedule for $p$ extends to $p\oplus v$ without increasing any load by assigning the padding job to a machine on which it has zero processing time. Conversely, deleting the padding job from any schedule for $p\oplus v$ cannot increase its makespan. Hence
\begin{equation*}
\operatorname{OPT}(p\oplus v)=\operatorname{OPT}(p)>0~.
\end{equation*}
The transformed allocation keeps only the original jobs, so
\begin{equation*}
\begin{aligned}
C\bigl(p,[T_v(x)]^S(p)\bigr)
&=\max_{i\in[m]}\sum_{j\in S}p_{ij}x_{ij}^{S\cup\{k\}}(p\oplus v)\\
&\le\max_{i\in[m]}\left(
\sum_{j\in S}p_{ij}x_{ij}^{S\cup\{k\}}(p\oplus v)
+v_i x_{ik}^{S\cup\{k\}}(p\oplus v)\right)\\
&=C\bigl(p\oplus v,x^{S\cup\{k\}}(p\oplus v)\bigr)\\
&\le\Gamma\operatorname{OPT}(p\oplus v)
=\Gamma\operatorname{OPT}(p)~.
\end{aligned}
\end{equation*}
The final inequality applies the original approximation guarantee to the enlarged instance, whose optimum is positive. Thus $T_v(x)$ preserves the same approximation guarantee $\Gamma$.
\end{proof}

One application of $T_v$ need not produce an invariant mechanism. Intuitively, we would like to apply padding infinitely often over all possible padding jobs and obtain a limiting mechanism that is unchanged by any further padding. Since each operation preserves truthfulness and the approximation guarantee, such a limit is a natural candidate for the replacement we seek. The classical Markov--Kakutani fixed-point theorem, stated in Lemma~\ref{lem:padding-common-fixed-point}, provides a rigorous way to realize this intuition. In the proof of Lemma~\ref{lem:padding-invariant-family}, we verify that the padding-appending operations are commuting continuous affine self-maps of a nonempty compact convex class of mechanisms. They therefore have a common fixed point, which is precisely a mechanism invariant under every padding operation.

\begin{lemma}[Markov--Kakutani fixed-point theorem {\cite{kakutani1938two}}]
\label{lem:padding-common-fixed-point}
Let $K$ be a nonempty compact convex subset of a product of real coordinate spaces. Every commuting family of continuous affine maps from $K$ to itself has a common fixed point.
\end{lemma}

\begin{lemma}[Padding-invariant replacement]
\label{lem:padding-invariant-family}
Every truthful, $\Gamma$-approximate fractional mechanism family $(x,u)$ admits a truthful, $\Gamma$-approximate fractional replacement $(\widehat x,\widehat u)$ with finite implementing payments that is invariant under every padding-appending operation. Namely, for every $v\in\mathbb R_{\ge0}^m$ with $\min_i v_i=0$,
\begin{equation*}
T_v(\widehat x)=\widehat x,
\qquad T_v(\widehat u)=\widehat u~.
\end{equation*}
\end{lemma}

\begin{proof}
Let $K_\Gamma$ be the class of truthful, $\Gamma$-approximate fractional mechanism families that are equivariant under job relabeling, represented by their allocation rules and normalized utilities. We first show that $K_\Gamma$ is nonempty, compact, and convex. We then show that the padding-appending operations are continuous affine self-maps of $K_\Gamma$ and commute with one another. Applying Lemma~\ref{lem:padding-common-fixed-point} will give a common fixed point in $K_\Gamma$, yielding the desired replacement.

To show that $K_\Gamma$ is nonempty, we symmetrize the given mechanism over job labels. For a finite nonempty job set $S$ of size $n$ and a bijection $b:S\to[n]$, let $p^b$ be the relabeled report, with $p^b_{i,b(j)}=p_{ij}$. Starting from the given mechanism family, define
\begin{equation*}
\begin{aligned}
\bar x_{ij}^S(p)
&=\frac1{n!}\sum_{b:S\to[n]}x_{i,b(j)}^{[n]}(p^b),\\
\bar u_i^S(p)
&=\frac1{n!}\sum_{b:S\to[n]}u_i^{[n]}(p^b).
\end{aligned}
\quad\Longrightarrow\quad
\begin{aligned}
\bar u_i^S(p)
&\ge \bar u_i^S(q)+(q_i-p_i)\cdot\bar x_i^S(q),\\
\sum_{j\in S}p_{ij}\bar x_{ij}^S(p)
&=\frac1{n!}\sum_{b:S\to[n]}\sum_{\ell\in[n]}p^b_{i\ell}x_{i\ell}^{[n]}(p^b)\\
&\le\frac1{n!}\sum_{b:S\to[n]}\Gamma\operatorname{OPT}(p^b)
=\Gamma\operatorname{OPT}(p).
\end{aligned}
\end{equation*}
Here the first inequality holds for all nonnegative reports $p,q$ differing only in machine $i$'s row, and the load bound holds whenever $\operatorname{OPT}(p)>0$. Thus the matched averages preserve truthfulness and the approximation guarantee $\Gamma$, as well as feasibility and zero-row normalization. Relabeling the jobs merely reindexes the averages, so the resulting family is equivariant under job relabeling. Using the unique empty allocation and zero utilities on the empty job set gives an element of $K_\Gamma$.

For compactness, feasibility and Lemma~\ref{lem:utility-bounds} give
\begin{equation*}
x_{ij}^S(p)\in[0,1],
\qquad u_i^S(p)\in\left[-\sum_{j\in S}p_{ij},\,0\right].
\end{equation*}
The product of these compact intervals is compact. All defining constraints of $K_\Gamma$ are closed linear equalities or inequalities involving finitely many evaluation coordinates. Hence $K_\Gamma$ is closed in this product and therefore compact.

To establish convexity, observe that matched convex combinations of allocation and utility families preserve feasibility, normalization, the incentive inequalities, the $\Gamma$ load bounds, and equivariance under job relabeling. Thus
\begin{equation*}
\begin{gathered}
(x^{(1)},u^{(1)}),(x^{(2)},u^{(2)})\in K_\Gamma\;
\Longrightarrow\;
\lambda(x^{(1)},u^{(1)})+(1-\lambda)(x^{(2)},u^{(2)})\in K_\Gamma
\qquad(0\le\lambda\le1),
\end{gathered}
\end{equation*}
proving convexity.

We next verify that the padding-appending operations are continuous affine self-maps of $K_\Gamma$ and commute with one another. Consider the family
\begin{equation*}
\mathcal T=\{T_v:v\in\mathbb R_{\ge0}^m,\ \min_i v_i=0\}.
\end{equation*}
Equivariance under job relabeling makes each operation independent of the fresh-label choice and is preserved by padding. Together with Lemma~\ref{lem:padding-preservation} and the utility bounds from Lemma~\ref{lem:utility-bounds}, this gives $T_v(K_\Gamma)\subseteq K_\Gamma$. For each fixed evaluation coordinate and a fresh label $k$,
\begin{equation*}
\begin{aligned}
[T_v(x)]_{ij}^S(p)
&=x_{ij}^{S\cup\{k\}}(p\oplus v),\\
[T_v(u)]_i^S(p)
&=u_i^{S\cup\{k\}}(p\oplus v)
-u_i^{S\cup\{k\}}(p^{i,0}\oplus v).
\end{aligned}
\end{equation*}
Each output coordinate is a finite linear combination of input evaluation coordinates. Hence $T_v$ is continuous in the product topology and affine.

To verify commutativity, fix $T_v,T_w\in\mathcal T$ and use distinct fresh labels $k,\ell$ for the two padding jobs. Equivariance allows both composition orders to be evaluated on the same enlarged report $p\oplus v\oplus w$. For every core job $j\in S$,
\begin{equation*}
[T_v(T_w(x))]_{ij}^S(p)
=x_{ij}^{S\cup\{k,\ell\}}(p\oplus v\oplus w)
=[T_w(T_v(x))]_{ij}^S(p).
\end{equation*}
For utilities, fix machine $i$ and all opponents' reports on the enlarged job set. Write $H(r,a,b)$ for its utility when its core row is $r$ and its costs on jobs $k,\ell$ are $a,b$, respectively. Then
\begin{equation*}
\begin{aligned}
[T_v(T_w(u))]_i^S(p)
&=[H(p_i,v_i,w_i)-H(0,0,w_i)]-[H(0,v_i,w_i)-H(0,0,w_i)]\\
&=H(p_i,v_i,w_i)-H(0,v_i,w_i),\\[4pt]
[T_w(T_v(u))]_i^S(p)
&=[H(p_i,v_i,w_i)-H(0,v_i,0)]-[H(0,v_i,w_i)-H(0,v_i,0)]\\
&=H(p_i,v_i,w_i)-H(0,v_i,w_i).
\end{aligned}
\end{equation*}
Consequently, $T_vT_w=T_wT_v$ on $K_\Gamma$.

All hypotheses of Lemma~\ref{lem:padding-common-fixed-point} are now satisfied, so we obtain a common fixed point $(\widehat x,\widehat u)\in K_\Gamma$, satisfying
\begin{equation*}
T_v(\widehat x)=\widehat x,
\qquad T_v(\widehat u)=\widehat u
\qquad\text{for every }T_v\in\mathcal T.
\end{equation*}
Membership in $K_\Gamma$ preserves feasibility, normalization, the incentive inequalities, and the approximation guarantee $\Gamma$. The finite payments
\begin{equation*}
\widehat\pi_i^S(p)=\widehat u_i^S(p)+p_i\cdot\widehat x_i^S(p)
\end{equation*}
implement $\widehat x$ truthfully, completing the construction.
\end{proof}

Thus far, we have proved the existence of a truthful, $\Gamma$-approximate mechanism family that is padding-invariant. We write $(x,u)$ for this family and now study its normalized utilities $u_i^S$ and utility interactions $V_i^J$. The next lemma shows that these interactions satisfy the zero boundary condition introduced in Section~\ref{sec:utility-interactions}.

\begin{lemma}[Zero boundary condition]
\label{lem:utility-zero-boundary}
Let $(x,u)$ be a padding-invariant truthful fractional mechanism family. For every finite job set $J$, every nonnegative report $p_J$, and every machine $i$,
$$
p_{ik}=0\text{ for some }k\in J
\quad\Longrightarrow\quad V_i^J(p_J)=0~.
$$
Here $V_i^J$ is the utility interaction of $(x,u)$.
\end{lemma}

\begin{proof}
Fix $J$, $p_J$, $i$, and $k\in J$ with $p_{ik}=0$. Let $v=(p_{ak})_{a\in[m]}$ be the processing-time column of job $k$. Since $v_i=0$, this is a valid padding job.

For each $S\subseteq J$ containing $k$, view $k$ as padding around $S\setminus\{k\}$. The reference report sets machine $i$'s costs on $S\setminus\{k\}$ to zero and retains its cost $v_i=p_{ik}=0$ on job $k$. Hence
\begin{equation*}
(p_{S\setminus\{k\}})^{i,0}\oplus v
=(p_S)^{i,0},
\qquad u_i^S\bigl((p_S)^{i,0}\bigr)=0.
\end{equation*}
The padding fixed-point identity therefore gives
\begin{equation*}
\begin{aligned}
u_i^{S\setminus\{k\}}(p_{S\setminus\{k\}})=[T_v(u)]_i^{S\setminus\{k\}}(p_{S\setminus\{k\}})&=u_i^S(p_S)
-u_i^S\bigl((p_{S\setminus\{k\}})^{i,0}\oplus v\bigr)\\
&=u_i^S(p_S)
-\underbrace{u_i^S\bigl((p_S)^{i,0}\bigr)}_{=0}=u_i^S(p_S).
\end{aligned}
\end{equation*}
Pairing each subset containing $k$ with the subset obtained by deleting $k$, we obtain
\begin{equation*}
\begin{aligned}
V_i^J(p_J)
&=\sum_{S\subseteq J}(-1)^{|J|-|S|}u_i^S(p_S)=\sum_{\substack{S\subseteq J\\k\in S}}
(-1)^{|J|-|S|}
\left(u_i^S(p_S)
-u_i^{S\setminus\{k\}}(p_{S\setminus\{k\}})\right)=0.
\end{aligned}
\end{equation*}
This includes the pair $S=\{k\}$ and $S\setminus\{k\}=\varnothing$, with $u_i^\varnothing=0$.
\end{proof}

\subsection{Vanishing Utility Interactions}
\label{sec:interaction-boundary-smoothing}
In this subsection, we aim to prove that $V_i^J=0$ for $|J|\ge m$ on strictly positive reports using two properties: (1) the zero boundary condition and (2) the differential conservation law. The padding-invariant mechanism constructed in Section~\ref{sec:padding-replacement} already satisfies (1). If we additionally assume that its allocation rules are continuous on strictly positive reports, then Lemma~\ref{lem:utility-interaction-conservation} gives (2). We temporarily adopt this assumption and show how the two properties force the utility interactions to vanish. In Section~\ref{sec:interaction-regularization}, we will remove the continuity assumption by smoothing.

Our plan is to apply the multivariate Laplace transform to each utility interaction $V_i^J$ and prove that its transform is zero for every positive weight matrix $z$. The uniqueness theorem for the Laplace transform states that, for continuous functions of at most linear growth, equality of their transforms at every positive $z$ implies equality of the functions. Since the zero function has zero Laplace transform, this will imply $V_i^J=0$. We state the analytic step separately so that the same argument can later be applied to smoothed interactions.


\begin{lemma}[Vanishing via Laplace transform]
\label{lem:laplace-annihilation}
Let $J$ be a finite job set with $|J|\ge m$, and let $D=m|J|$.
For each $i\in[m]$, let $f_i^J:[0,\infty)^D\to\mathbb R$
be differentiable along each own-coordinate line whose other
coordinates are strictly positive, with a right derivative at
the zero endpoint. Assume the following:
\begin{enumerate}[label=(\arabic*)]
\item \emph{Continuity.}
Each $f_i^J$ is continuous on $(0,\infty)^D$.

\item \emph{Bounded own-coordinate derivatives.}
There exists $B<\infty$ such that
\[
\left|\frac{\partial f_i^J(p)}{\partial p_{ij}}\right|\le B
\qquad
\forall p\in(0,\infty)^D,\ i\in[m],\ j\in J~.
\]

\item \emph{Zero boundary condition.}
For every $i\in[m]$, $j\in J$, and $p\in[0,\infty)^D$,
\[
p_{ij}=0\quad\Longrightarrow\quad f_i^J(p)=0~.
\]

\item \emph{Differential conservation law.}
For every $p\in(0,\infty)^D$,
\begin{equation}
\sum_{i=1}^m\frac{\partial f_i^J(p)}{\partial p_{ij}}=0
\qquad\text{for every }j\in J~.
\label{eq:laplace-differential-system}
\end{equation}
\end{enumerate}
Then
\[
f_i^J(p)=0
\qquad
\forall i\in[m],\ \forall p\in(0,\infty)^D~.
\]
\end{lemma}

\begin{proof}
We first apply the Laplace transform to convert the differential
identities into a linear system. We then prove the lemma by contradiction, using the uniqueness
and continuity of the transforms and the density of
full-column-rank matrices.

To justify the transforms and the subsequent integration steps,
we begin with a growth bound. Fix $i\in[m]$ and $j\in J$, and
hold all coordinates other than $p_{ij}$ at strictly positive
values. The differentiability assumptions give continuity along
this line, including at zero. The mean value theorem and
condition~(2) therefore imply that $f_i^J$ is $B$-Lipschitz in
$p_{ij}$ on $[0,\infty)$, and hence absolutely continuous on
every bounded interval. Together with condition~(3), this gives
\[
\bigl|f_i^J(p)\bigr|
=
\bigl|f_i^J(p)-f_i^J(p|_{p_{ij}=0})\bigr|
\le Bp_{ij}~.
\]
Thus each $f_i^J$ has at most linear growth. For
$z\in(0,\infty)^{m\times J}$, define
\begin{equation*}
L_i^J(z)
=
\int_{(0,\infty)^D}
e^{-\langle z,p\rangle}f_i^J(p)\,dp~,
\qquad
\langle z,p\rangle
=
\sum_{k\in[m],\,j\in J}z_{kj}p_{kj}~.
\end{equation*}
The growth and derivative bounds imply that both $f_i^J$ and
its own-coordinate derivatives are absolutely integrable after
multiplication by $e^{-\langle z,p\rangle}$. Fubini's theorem
therefore allows us to integrate first in $p_{ij}$ and then in
the remaining coordinates.

We next use integration by parts to express the transforms of
the derivatives in terms of $L_i^J$. Fix all coordinates other
than $p_{ij}$ at positive values, and let $p(t)$ denote the
report with $p_{ij}$ replaced by $t$. Then
\[
\begin{aligned}
\int_0^\infty
e^{-z_{ij}t}\frac{d}{dt}f_i^J(p(t))\,dt
&=
\bigl[e^{-z_{ij}t}f_i^J(p(t))\bigr]_{0}^{\infty}
+
z_{ij}\int_0^\infty
e^{-z_{ij}t}f_i^J(p(t))\,dt\\
&=
z_{ij}\int_0^\infty
e^{-z_{ij}t}f_i^J(p(t))\,dt~.
\end{aligned}
\]
The boundary term at zero is zero by condition~(3), while
$e^{-z_{ij}t}f_i^J(p(t))\to0$ as $t\to\infty$ because
$|f_i^J(p(t))|\le Bt$ and $z_{ij}>0$.
Multiplying by the exponential weights for the remaining
coordinates and integrating gives
\[
\int_{(0,\infty)^D}
e^{-\langle z,p\rangle}
\frac{\partial f_i^J(p)}{\partial p_{ij}}\,dp
=
z_{ij}L_i^J(z)~.
\]
Applying this identity to
\eqref{eq:laplace-differential-system}, we obtain
\begin{equation*}
0
=
\int_{(0,\infty)^D}
e^{-\langle z,p\rangle}
\sum_{i=1}^m
\frac{\partial f_i^J(p)}{\partial p_{ij}}\,dp
=
\sum_{i=1}^m z_{ij}L_i^J(z)
\qquad(j\in J)~.
\end{equation*}
Writing $J=\{j_1,\ldots,j_n\}$, where $n\ge m$, we express
these equations in matrix form as
\begin{equation}
\underbrace{
\begin{pmatrix}
z_{1j_1}&\cdots&z_{mj_1}\\
\vdots&&\vdots\\
z_{1j_n}&\cdots&z_{mj_n}
\end{pmatrix}
}_{Z(z)}
\begin{pmatrix}
L_1^J(z)\\
\vdots\\
L_m^J(z)
\end{pmatrix}
=0~.
\label{eq:laplace-matrix-system}
\end{equation}
This system holds for every $z\in(0,\infty)^{m\times J}$.

We now argue by contradiction. Suppose that
$f_{i_0}^J(p^0)\neq0$ for some $i_0\in[m]$ and
$p^0\in(0,\infty)^D$. By continuity, $f_{i_0}^J$ is nonzero
throughout a neighborhood of $p^0$, so it is not zero almost
everywhere. The uniqueness theorem for multivariate Laplace
transforms~\cite[Theorem~E]{kuvcera1968multiple} therefore
implies that $L_{i_0}^J$ cannot be identically zero on the
positive parameter domain. Hence, there exists
$z^0\in(0,\infty)^{m\times J}$ such that
\[
L_{i_0}^J(z^0)\neq0~.
\]
The growth bound also ensures absolute convergence of the
transform whenever every coordinate of $z$ has positive real
part. By analyticity in the interior of the convergence
domain~\cite[Theorem~D]{kuvcera1968multiple},
$L_{i_0}^J$ is therefore continuous on
$(0,\infty)^{m\times J}$. Consequently, there is an open
neighborhood $U$ of $z^0$, contained in the positive parameter
domain, such that
\[
L_{i_0}^J(z)\neq0
\qquad\text{for every }z\in U~.
\]

Since $n\ge m$, matrices with full column rank are dense in
$\mathbb R^{n\times m}$~\cite[Section~2.5.5, p.~73]{DBLP:books/daglib/0086372}.
Thus $U$ contains a parameter choice $z^*$ for which $Z(z^*)$
has full column rank. Applying
\eqref{eq:laplace-matrix-system} at $z=z^*$ forces
$L_i^J(z^*)=0$ for every $i\in[m]$. In particular,
$L_{i_0}^J(z^*)=0$, contradicting the defining property of $U$.
Therefore, we have
\[
f_i^J(p)=0
\qquad
\forall i\in[m],\ \forall p\in(0,\infty)^D~,
\]
completing the proof.
\end{proof}

\begin{lemma}[Vanishing interactions under continuity]
\label{lem:vanishing-differentiable-interactions}
Suppose that the padding-invariant allocation rules $x^S$ are continuous on strictly positive reports. If $|J|\ge m$, then $V_i^J(p)=G_{ij}^J(p)=0$ for every strictly positive report $p$, every $i\in[m]$, and every $j\in J$.
\end{lemma}

\begin{proof}
We apply Lemma~\ref{lem:laplace-annihilation} to the utility interactions $V_i^J$ of the padding-invariant mechanism. By truthfulness and the difference-quotient bounds derived from \eqref{eq:normalized-ic}, each $u_i^S$ is convex and $1$-Lipschitz in every own coordinate, so its right derivative at zero exists and is finite. The same endpoint differentiability holds for the finite linear combination $V_i^J$. We now verify the four conditions of Lemma~\ref{lem:laplace-annihilation}.

\begin{enumerate}[label=(\arabic*)]
\item \emph{Continuity.} By Lemma~\ref{lem:utility-derivatives}, $\partial u_i^S/\partial p_{ij}=-x_{ij}^S$ on strictly positive reports. Integrating along the own-row segment from zero to $p_i$, with the endpoint justified by normalization and the utility bounds in Lemma~\ref{lem:utility-bounds}, gives
$$
u_i^S(p)=-\int_0^1 p_i\cdot x_i^S(tp_i,p_{-i})\,dt~.
$$
Continuity of $x^S$ on strictly positive reports and boundedness of allocations imply, by dominated convergence, that $u_i^S$ is jointly continuous there. Hence its finite linear combination $V_i^J$ satisfies condition~(1).

\item \emph{Bounded own-coordinate derivatives.} By Lemma~\ref{lem:utility-derivatives} and $0\le x_{ij}^S\le1$,
$$
\left|\frac{\partial V_i^J(p)}{\partial p_{ij}}\right|
=|G_{ij}^J(p)|
=\left|\sum_{\substack{S\subseteq J\\j\in S}}
(-1)^{|J|-|S|}x_{ij}^S(p_S)\right|
\le\sum_{\substack{S\subseteq J\\j\in S}}x_{ij}^S(p_S)
\le 2^{|J|-1}~.
$$
Thus condition~(2) holds with $B=2^{|J|-1}$.

\item \emph{Zero boundary condition.} Lemma~\ref{lem:utility-zero-boundary} gives $V_i^J(p)=0$ whenever $p_{ik}=0$ for some $k\in J$, establishing condition~(3).
\item \emph{Differential conservation law.} Since $|J|\ge m\ge2$, Lemma~\ref{lem:utility-interaction-conservation} gives $\sum_{i=1}^m\partial V_i^J(p)/\partial p_{ij}=0$ for every $j\in J$ and strictly positive $p$, establishing condition~(4).
\end{enumerate}

Applying Lemma~\ref{lem:laplace-annihilation} with $f_i^J=V_i^J$, we conclude that $V_i^J=0$ on strictly positive reports. Consequently,
$$
G_{ij}^J(p)=-\frac{\partial V_i^J(p)}{\partial p_{ij}}=0~,
$$
which proves the lemma.
\end{proof}

We next remove the continuity assumption and construct one replacement whose interaction identities also hold on boundary reports.

\subsection{Removing the Continuity Assumption}
\label{sec:interaction-regularization}
\label{sec:final-interaction-replacement}
\label{sec:vanishing-smoothed-interactions}

We return to the padding-invariant mechanism without assuming continuity of its allocation rules. Our goal is to construct a smooth truthful family $(x^\delta,u^\delta)$ whose utility interactions $V_i^{J,\delta}$ vanish whenever $|J|\ge m$. A natural approach is the standard right-smoothing operation: at each report $p$, replace a function's value by a weighted average of its values at nearby reports $p+r$, where each coordinate of $r$ lies in a small positive interval. Thus, we average over a small box to the right of $p$ in every coordinate. Applying this operation to $x$ and $u$, followed by utility normalization, gives a smooth truthful family. However, its zero boundary condition is not immediate: even when $p_{ij}=0$, the average samples reports with $p_{ij}+r_{ij}>0$. We therefore cannot yet apply Lemma~\ref{lem:laplace-annihilation} to the new utility interactions.

To overcome this difficulty, we first extend the original utility interactions $V_i^J$ by zero and study their left convolutions $V_i^J\ast_{\mathrm L}\kappa$. Left smoothing instead averages nearby values at $p-r$, in a small box to the left of $p$ in every coordinate. At a zero own coordinate, these reports lie outside the nonnegative domain, where the zero extension makes every sampled value zero. These left-smoothed interactions are smooth and have zero boundary values. We verify the conditions in Lemma~\ref{lem:laplace-annihilation}. The lemma then gives $V_i^J\ast_{\mathrm L}\kappa=0$ whenever $|J|\ge m$.

The key step is to identify the right smoothing of $V_i^J$ with its left smoothing at a shifted report, using the symmetry of the product kernel. Since the left smoothing is zero everywhere, the right smoothing is zero as well. Consequently, the new mechanism satisfies $V_i^{J,\delta}=0$ for every nonnegative report and $|J|\ge m$, so its allocation rule has no interactions of order at least $m$. We first analyze the left-smoothed interactions, then carry out the right smoothing and verify this correspondence.

We write $\ast_{\mathrm L}$ and $\ast_{\mathrm R}$ for the two averaging operations
$$
\begin{aligned}
(h\ast_{\mathrm L}\kappa)(p)
&:=\int_{\mathbb R^D}h(p-r)\kappa(r)\,dr,\\
(h\ast_{\mathrm R}\kappa)(p)
&:=\int_{\mathbb R^D}h(p+r)\kappa(r)\,dr.
\end{aligned}
$$
The left operation is ordinary convolution. When $\kappa$ is supported in the positive orthant, these operations average reports to the left and right of $p$, respectively, coordinatewise.

\paragraph{Zero extension and left smoothing.}
Fix a job set $J$ with $n=|J|\ge2$, and set $D=mn$. The construction has two steps:
\begin{enumerate}[label=Step \arabic*.,leftmargin=*]
\item \emph{Zero extension.} Extend $V_i^J$ and $G_{ij}^J$ to $\mathbb R^D$ by keeping their values on $[0,\infty)^D$ and setting them to zero whenever any coordinate is negative. We use the same symbols for these extensions.
\item \emph{Left smoothing.} Choose a nonnegative smooth kernel $\kappa:\mathbb R^D\to\mathbb R$ with compact support contained in $(0,\infty)^D$ and integral one. Here smooth means infinitely differentiable. Lemma~\ref{lem:smooth-kernels} below provides such a kernel. We study the ordinary convolutions $V_i^J\ast_{\mathrm L}\kappa$, which average reports $p-r$ with $r>0$.\footnote{\label{fn:measurability}Here and below, writing these averages as ordinary Lebesgue integrals implicitly assumes joint measurability of the allocation rules and normalized utilities. Appendix~\ref{app:nonmeasurable-rules} removes this assumption without changing the conclusions.}
\end{enumerate}

\paragraph{Justifying the construction.}
To show that the left-smoothed interactions are well-defined, we need to establish that the kernel $\kappa$ exists and that $V_i^J$ is bounded on every bounded region, so the convolution integral is finite. We prove these two facts first, then establish smoothness and verify the zero boundary condition and the differential conservation law.

\begin{lemma}
\label{lem:smooth-kernels}
For every $\delta>0$, there is a nonnegative infinitely differentiable function $k_\delta:\mathbb R\to\mathbb R$ that is zero outside $[\delta/4,3\delta/4]$ and satisfies $k_\delta(t)=k_\delta(\delta-t)$ and $\int_{\mathbb R}k_\delta(t)\,dt=1$. For every positive integer $D$, the product kernel
$$
\kappa(r)=\prod_{\ell=1}^D k_\delta(r_\ell)
$$
is nonnegative, infinitely differentiable, zero outside $[\delta/4,3\delta/4]^D$, and has integral one.
\end{lemma}

\begin{proof}
Start with the standard smooth bump function
$$
\rho(s)=
\begin{cases}
\exp\!\left(-\dfrac{1}{1-s^2}\right),& |s|<1,\\
0,& |s|\ge1.
\end{cases}
$$
This function is nonnegative and positive on $(-1,1)$. All its derivatives approach zero at $s=\pm1$, so $\rho\in C^\infty(\mathbb R)$. Let $C=\int_{\mathbb R}\rho(s)\,ds$, where $0<C<\infty$, and define
$$
k_\delta(t)=\frac{4}{C\delta}\,\rho\!\left(\frac{4t}{\delta}-2\right).
$$
Since $\rho$ is even, $k_\delta(\delta-t)=k_\delta(t)$. Also, $k_\delta$ is nonnegative and infinitely differentiable. Its support is contained in $[\delta/4,3\delta/4]$, since
$$
\left|\frac{4t}{\delta}-2\right|<1
\quad\Longleftrightarrow\quad
\frac{\delta}{4}<t<\frac{3\delta}{4}.
$$
The change of variables $s=4t/\delta-2$ gives
$$
\int_{\mathbb R}k_\delta(t)\,dt
=\frac{4}{C\delta}\cdot\frac{\delta}{4}
\int_{\mathbb R}\rho(s)\,ds=1.
$$
For any positive integer $D$, the product $\kappa(r)=\prod_{\ell=1}^D k_\delta(r_\ell)$ is therefore nonnegative, infinitely differentiable, and supported in $[\delta/4,3\delta/4]^D$. Finally, iterated integration gives
$$
\int_{\mathbb R^D}\kappa(r)\,dr
=\prod_{\ell=1}^D\int_{\mathbb R}k_\delta(r_\ell)\,dr_\ell=1.
$$
\end{proof}

We first record bounds ensuring that the convolution integrals are well-defined. Set $N=2^{|J|-1}$. The utility bounds and the definitions of the interactions give, for every $p\in[0,\infty)^D$,
\begin{equation*}
|V_i^J(p)|\le N\sum_{j\in J}p_{ij},
\qquad |G_{ij}^J(p)|\le N~,
\end{equation*}
These bounds also show that the zero-extended interactions are bounded on every bounded region of $\mathbb R^D$. For each fixed $p$, the compact support of $\kappa$ restricts the integral
$$
(V_i^J\ast_{\mathrm L}\kappa)(p)
=\int_{\mathbb R^D}V_i^J(s)\kappa(p-s)\,ds
$$
to such a region. Hence the integral is absolutely convergent.

Writing $(V_i^J\ast_{\mathrm L}\kappa)(p)=\int_{\mathbb R^D}V_i^J(s)\kappa(p-s)\,ds$, the dependence on $p$ is entirely through the smooth kernel $\kappa(p-s)$. When $p$ ranges over a compact set, the kernel and all its derivatives vanish outside a common bounded region of $s$, where $V_i^J$ is bounded. Dominated convergence therefore allows us to differentiate under the integral any number of times, and the resulting derivatives are continuous. Thus $V_i^J\ast_{\mathrm L}\kappa$ is smooth on $\mathbb R^D$.

We next establish the derivative identity for the smoothed interactions. The following identity uses ordinary convolution $\ast_{\mathrm L}$ and holds for an arbitrary nonnegative smooth compactly supported kernel $\kappa$, without a one-sided support or symmetry requirement.

\begin{lemma}
\label{lem:smoothed-utility-derivative}
Let $(x,u)$ be a padding-invariant truthful fractional mechanism family, fix $|J|\ge2$, and set $D=m|J|$. For its zero-extended interactions and every nonnegative smooth compactly supported kernel $\kappa:\mathbb R^D\to\mathbb R$, the following identities hold on $\mathbb R^D$:
\begin{equation}
\frac{\partial}{\partial p_{ij}}(V_i^J\ast_{\mathrm L}\kappa)
=-(G_{ij}^J\ast_{\mathrm L}\kappa),
\qquad
\sum_{i=1}^m\frac{\partial}{\partial p_{ij}}(V_i^J\ast_{\mathrm L}\kappa)=0
\label{eq:smoothed-utility-derivative}
\end{equation}
for every $i\in[m]$ and $j\in J$.
\end{lemma}

\begin{proof}
Suppose that the zero-extended $V_i^J$ is Lipschitz along each own-coordinate line and satisfies $\partial V_i^J/\partial s_{ij}=-G_{ij}^J$ almost everywhere along that line. Lipschitz continuity implies absolute continuity on bounded intervals, allowing integration by parts. Thus,
$$
\begin{aligned}
\frac{\partial}{\partial p_{ij}}(V_i^J\ast_{\mathrm L}\kappa)(p)
&=\int_{\mathbb R^D}V_i^J(s)\frac{\partial}{\partial p_{ij}}\kappa(p-s)\,ds\\
&=-\int_{\mathbb R^D}V_i^J(s)\frac{\partial}{\partial s_{ij}}\kappa(p-s)\,ds\\
&=\int_{\mathbb R^D}\frac{\partial V_i^J(s)}{\partial s_{ij}}\kappa(p-s)\,ds\\
&=-\int_{\mathbb R^D}G_{ij}^J(s)\kappa(p-s)\,ds\\
&=-(G_{ij}^J\ast_{\mathrm L}\kappa)(p).
\end{aligned}
$$
The smooth compact support of $\kappa$ and the local bounds on the interactions justify differentiation and the iterated integrals, and eliminate boundary terms at infinity. Complete allocation gives $\sum_iG_{ij}^J=0$ on nonnegative reports, and zero extension preserves this identity on $\mathbb R^D$. Hence
$$
\sum_{i=1}^m\frac{\partial}{\partial p_{ij}}(V_i^J\ast_{\mathrm L}\kappa)
=-\left(\sum_{i=1}^mG_{ij}^J\right)\ast_{\mathrm L}\kappa=0.
$$

It remains to verify the two properties used above. Fix an own coordinate $s_{ij}$ and write $s(t)$ for the report obtained by setting $s_{ij}=t$, with all other coordinates held fixed.

We first establish Lipschitz continuity along own coordinates by showing that, along every own-coordinate line $s(t)$, the zero-extended $V_i^J$ satisfies the bound
$$
|V_i^J(s(b))-V_i^J(s(a))|\le N|b-a|
\qquad\text{for all }a,b\in\mathbb R,
\quad N=2^{|J|-1}.
$$
Suppose first that all other fixed coordinates are nonnegative. For $a,b\ge0$, the difference-quotient bounds derived from \eqref{eq:normalized-ic} and $0\le x_{ij}^S\le1$ give $|u_i^S(s(b)_S)-u_i^S(s(a)_S)|\le|b-a|$ whenever $j\in S$. Terms with $j\notin S$ do not change, so
$$
\begin{aligned}
|V_i^J(s(b))-V_i^J(s(a))|
&\le\sum_{\substack{S\subseteq J\\j\in S}}
|u_i^S(s(b)_S)-u_i^S(s(a)_S)|\\
&\le 2^{|J|-1}|b-a|.
\end{aligned}
$$
Lemma~\ref{lem:utility-zero-boundary} gives $V_i^J(s(0))=0$. Thus, if $a<0\le b$, zero extension gives
$$
|V_i^J(s(b))-V_i^J(s(a))|
=|V_i^J(s(b))-V_i^J(s(0))|
\le Nb\le N|b-a|.
$$
The case $b<0\le a$ follows by symmetry. If both parameters are negative, or any other fixed coordinate is negative, the extended function is zero at both points. This proves the claimed bound on the entire line.

We next verify the almost-everywhere derivative identity. Fix all other coordinates at nonnegative values. For each $S\subseteq J$ containing $j$, the difference-quotient bounds give, for $h>0$,
$$
-x_{ij}^S(s(t)_S)
\le\frac{u_i^S(s(t+h)_S)-u_i^S(s(t)_S)}h
\le-x_{ij}^S(s(t+h)_S).
$$
In particular, $x_{ij}^S(s(t)_S)\ge x_{ij}^S(s(t+h)_S)$, so this allocation is nonincreasing in $t$. A monotone real-valued function has at most countably many discontinuities. At every positive continuity point, letting $h\downarrow0$ in the displayed bounds, and similarly in the bounds for the interval from $t-h$ to $t$, gives
$$
\frac{d}{dt}u_i^S(s(t)_S)=-x_{ij}^S(s(t)_S).
$$
There are finitely many relevant subsets $S$, so the union of their discontinuity sets is still countable and hence has measure zero. Outside this union, all these derivatives exist simultaneously. Terms with $j\notin S$ are independent of $t$, and therefore
$$
\begin{aligned}
\frac{d}{dt}V_i^J(s(t))
&=-\sum_{\substack{S\subseteq J\\j\in S}}
(-1)^{|J|-|S|}x_{ij}^S(s(t)_S)\\
&=-G_{ij}^J(s(t))
\qquad\text{for almost every }t>0.
\end{aligned}
$$
For $t<0$, both extended interactions vanish. If any other fixed coordinate is negative, they vanish along the entire line. Hence the derivative identity holds almost everywhere on every own-coordinate line.
\end{proof}

We can now verify all four conditions of Lemma~\ref{lem:laplace-annihilation} for the left-smoothed interactions.

\begin{lemma}
\label{lem:vanishing-auxiliary-functions}
Let $(x,u)$ be a padding-invariant truthful fractional mechanism family. Fix $|J|\ge m$, set $D=m|J|$, and let $\kappa:\mathbb R^D\to\mathbb R$ be nonnegative and smooth, with integral one and compact support contained in $(0,\infty)^D$. For the zero-extended utility interactions of $(x,u)$, the left convolutions $V_i^J\ast_{\mathrm L}\kappa$ satisfy all the hypotheses of Lemma~\ref{lem:laplace-annihilation}. Consequently,
$$
(V_i^J\ast_{\mathrm L}\kappa)(p)=0\qquad\forall i\in[m],\ \forall p\in\mathbb R^D~.
$$
\end{lemma}

\begin{proof}
We check the conditions in the order stated in Lemma~\ref{lem:laplace-annihilation}.
\begin{enumerate}[label=(\arabic*)]
\item \emph{Continuity.} The smoothing construction makes every $V_i^J\ast_{\mathrm L}\kappa$ smooth on $\mathbb R^D$, and hence continuous on $(0,\infty)^D$.
\item \emph{Bounded own-coordinate derivatives.} Smoothness gives the required derivatives on the entire domain. Lemma~\ref{lem:smoothed-utility-derivative}, the bound $|G_{ij}^J|\le N$, positivity of integration, and the normalization of $\kappa$ give
$$
\left|\frac{\partial}{\partial p_{ij}}(V_i^J\ast_{\mathrm L}\kappa)(p)\right|
=|(G_{ij}^J\ast_{\mathrm L}\kappa)(p)|\le N~.
$$
\item \emph{Zero boundary condition.} If $p_{ij}=0$ and $\kappa(p-s)\ne0$, then $s_{ij}<0$, so the zero extension satisfies $V_i^J(s)=0$. Thus every term in the average defining $(V_i^J\ast_{\mathrm L}\kappa)(p)$ is zero. The same argument applies whenever any coordinate of $p$ is nonpositive, so every $V_i^J\ast_{\mathrm L}\kappa$ is zero outside the strictly positive orthant as well.
\item \emph{Differential conservation law.} Lemma~\ref{lem:smoothed-utility-derivative} directly gives $\sum_{i=1}^m\partial(V_i^J\ast_{\mathrm L}\kappa)/\partial p_{ij}=0$ for every $j\in J$, establishing condition~(4).
\end{enumerate}
Lemma~\ref{lem:laplace-annihilation} now gives $V_i^J\ast_{\mathrm L}\kappa=0$ on $(0,\infty)^D$. Together with the vanishing outside this orthant proved in (3), this establishes $V_i^J\ast_{\mathrm L}\kappa=0$ on all of $\mathbb R^D$.
\end{proof}

\paragraph{Realizing vanishing interactions with a truthful mechanism.}
We now carry out the right-smoothing construction proposed at the beginning of this subsection. We average the allocation rules $x$ and utilities $u$ to obtain a smooth replacement. The left-smoothed interactions have already been shown to vanish for $|J|\ge m$. Expressing the right averages as shifted left averages will therefore give the required vanishing interactions. The lemma below verifies this correspondence, truthfulness, and the stated additive approximation loss.

Fix $\delta>0$. To average allocations and utilities consistently across different job sets, choose one kernel $k_\delta$ from Lemma~\ref{lem:smooth-kernels} and, for every finite nonempty job set $S$, define
$$
\kappa_{S,\delta}(r)=\prod_{i\in[m],\,j\in S}k_\delta(r_{ij}).
$$
This kernel is nonnegative and smooth, has integral one, and is supported in $[\delta/4,3\delta/4]^{m\times S}$. Each factor is symmetric about $\delta/2$, so
$$
\kappa_{S,\delta}(\delta\mathbf1-r)=\kappa_{S,\delta}(r),
$$
where $\mathbf1$ is the all-ones matrix on $S$. Integrating out coordinates belonging to jobs outside a subset removes their normalized kernel factors and leaves exactly the kernel for that subset. This ensures compatibility with the finite sums defining utility interactions. For the empty set, use the unique empty allocation and zero utilities.

For a locally bounded function $h$ on nonnegative reports, extended by zero to $\mathbb R^{m\times S}$, right smoothing is
$$
\begin{aligned}
(h\ast_{\mathrm R}\kappa_{S,\delta})(p)
&=\int_{(0,\infty)^{m\times S}}h(p+r)\kappa_{S,\delta}(r)\,dr\\
&=\int_{\mathbb R^{m\times S}}h(s)\kappa_{S,\delta}(s-p)\,ds
\qquad(p\ge0).
\end{aligned}
$$
Thus the right average queries only strictly positive reports and is independent of the extension. The symmetry of the kernel gives the translation identity
\begin{equation}
(h\ast_{\mathrm R}\kappa_{S,\delta})(p)
=(h\ast_{\mathrm L}\kappa_{S,\delta})(p+\delta\mathbf1).
\label{eq:smoothing-translation}
\end{equation}
Indeed, $\kappa_{S,\delta}(s-p)=\kappa_{S,\delta}(p+\delta\mathbf1-s)$, so the two integrals have the same integrand.

We formalize this argument in the next lemma. It completes the task of this section by constructing a replacement satisfying the vanishing conclusion of Lemma~\ref{lem:vanishing-differentiable-interactions}, without any continuity assumption on the original allocation rules.

\begin{lemma}[Vanishing interactions without continuity]
\label{lem:truthful-smoothing}
Let $(x,u)$ be a truthful fractional mechanism family with approximation guarantee $\Gamma$. For every $\delta>0$, there exists a feasible truthful family $(x^\delta,u^\delta)$ whose allocation rules and normalized utilities are smooth and satisfy
\begin{equation}
C(p,x^{S,\delta}(p))\le\Gamma\bigl(\operatorname{OPT}(p)+|S|\delta\bigr)
\label{eq:regularized-performance}
\end{equation}
for every finite job set $S$ and every nonnegative report $p$ on $S$. If $(x,u)$ is padding-invariant, the new family also satisfies $G_{ij}^{J,\delta}(p)=0$ whenever $|J|\ge m$, for every $i\in[m]$, $j\in J$, and nonnegative report $p$ on $J$.
\end{lemma}

\begin{proof}
Fix $\delta>0$ and use the compatible product kernels $\kappa_{S,\delta}$ defined above.

We begin by constructing the new family. For every nonempty job set $S$ and nonnegative report $p$, set
\begin{equation}
\begin{aligned}
x_{ij}^{S,\delta}(p)
&=(x_{ij}^S\ast_{\mathrm R}\kappa_{S,\delta})(p),\\
u_i^{S,\delta}(p)
&=(u_i^S\ast_{\mathrm R}\kappa_{S,\delta})(p)
 -(u_i^S\ast_{\mathrm R}\kappa_{S,\delta})(p^{i,0}).
\end{aligned}
\label{eq:regularized-mechanism}
\end{equation}
These convolutions average the original functions at strictly positive reports $p+r$, with weight $\kappa_{S,\delta}(r)$. The subtracted reference utility ensures $u_i^{S,\delta}(p^{i,0})=0$. For the empty job set, use the empty allocation and zero utilities.

We now verify that these averages are smooth and feasible. Allocations are bounded, and the utility bounds in Lemma~\ref{lem:utility-bounds} imply that utilities are bounded on bounded regions. Thus the convolution integrals are well-defined. As in the left-smoothing argument, differentiating the smooth compactly supported kernel shows that each unnormalized average is smooth. The reference term is the averaged utility evaluated at $p^{i,0}$, so it is smooth as well. These functions are smooth also at boundary reports: the right average continues to query strictly positive reports in a neighborhood of every nonnegative input.

Nonnegativity of the kernel gives $x_{ij}^{S,\delta}(p)\ge0$. Complete allocation and normalization of the kernel give
$$
\begin{aligned}
\sum_{i=1}^m x_{ij}^{S,\delta}(p)
&=\int_{(0,\infty)^{m\times S}}
\left(\sum_{i=1}^m x_{ij}^S(p+r)\right)
\kappa_{S,\delta}(r)\,dr\\
&=\int_{(0,\infty)^{m\times S}}\kappa_{S,\delta}(r)\,dr=1.
\end{aligned}
$$
Hence the new allocation is feasible.

We next show that the averaging operation preserves truthfulness. Fix reports $p=(p_i,p_{-i})$ and $q=(q_i,p_{-i})$. For each perturbation $r$ in the averaging support, the original incentive inequality gives
$$
u_i^S(p+r)\ge u_i^S(q+r)+(q_i-p_i)\cdot x_i^S(q+r).
$$
The difference between the perturbed own rows remains $q_i-p_i$. Averaging this pointwise inequality against the same nonnegative kernel yields
$$
(u_i^S\ast_{\mathrm R}\kappa_{S,\delta})(p)
\ge (u_i^S\ast_{\mathrm R}\kappa_{S,\delta})(q)
+(q_i-p_i)\cdot(x_i^S\ast_{\mathrm R}\kappa_{S,\delta})(q).
$$
Since $p^{i,0}=q^{i,0}$, the reference utilities subtracted in the definitions of $u_i^{S,\delta}(p)$ and $u_i^{S,\delta}(q)$ are equal. Subtracting this common value gives
$$
u_i^{S,\delta}(p)\ge u_i^{S,\delta}(q)+(q_i-p_i)\cdot x_i^{S,\delta}(q).
$$
Together with zero-row normalization, this inequality implements the new allocations truthfully through the finite payments
$$
\pi_i^{S,\delta}(p)=u_i^{S,\delta}(p)+p_i\cdot x_i^{S,\delta}(p).
$$

To establish the approximation guarantee, observe that every perturbation with $\kappa_{S,\delta}(r)\ne0$ satisfies $0<r_{ij}<\delta$. Apply an optimal integral schedule for $p$ to $p+r$. Each machine receives at most $|S|$ jobs, each of whose processing times increases by less than $\delta$, so
$$
\operatorname{OPT}(p+r)\le\operatorname{OPT}(p)+|S|\delta.
$$
For each machine $i$, nonnegativity of allocations and the original approximation guarantee imply
$$
\begin{aligned}
\sum_{j\in S}p_{ij}x_{ij}^{S,\delta}(p)
&=\int_{(0,\infty)^{m\times S}}
\left(\sum_{j\in S}p_{ij}x_{ij}^S(p+r)\right)
\kappa_{S,\delta}(r)\,dr\\
&\le\int_{(0,\infty)^{m\times S}}
\left(\sum_{j\in S}(p_{ij}+r_{ij})x_{ij}^S(p+r)\right)
\kappa_{S,\delta}(r)\,dr\\
&\le\int_{(0,\infty)^{m\times S}}
\Gamma\operatorname{OPT}(p+r)\kappa_{S,\delta}(r)\,dr\\
&\le\Gamma\bigl(\operatorname{OPT}(p)+|S|\delta\bigr).
\end{aligned}
$$
Taking the maximum over machines proves the stated makespan bound, including when $\operatorname{OPT}(p)=0$.

Finally, assume that the original family is padding-invariant. We show that the new family has no interactions of order at least $m$. Fix $J$ with $|J|\ge m$. Lemma~\ref{lem:vanishing-auxiliary-functions} gives $V_i^J\ast_{\mathrm L}\kappa_{J,\delta}=0$ everywhere. Differentiating this zero function and applying Lemma~\ref{lem:smoothed-utility-derivative} also gives $G_{ij}^J\ast_{\mathrm L}\kappa_{J,\delta}=0$. The translation identity for the two averaging operations therefore yields
$$
\begin{aligned}
(V_i^J\ast_{\mathrm R}\kappa_{J,\delta})(p)
&=(V_i^J\ast_{\mathrm L}\kappa_{J,\delta})(p+\delta\mathbf1)=0,\\
(G_{ij}^J\ast_{\mathrm R}\kappa_{J,\delta})(p)
&=(G_{ij}^J\ast_{\mathrm L}\kappa_{J,\delta})(p+\delta\mathbf1)=0.
\end{aligned}
$$
We next identify these right convolutions with the interactions of the new mechanism. For every nonempty $S\subseteq J$, integrating out the coordinates belonging to $J\setminus S$ removes their normalized kernel factors. Thus, for any of the allocation or utility functions $h^S$ used below,
$$
(h^S\ast_{\mathrm R}\kappa_{S,\delta})(p_S)
=\int_{(0,\infty)^{m\times J}}
 h^S(p_S+r_S)\kappa_{J,\delta}(r)\,dr.
$$
Consequently, all subset averages can be written using the same kernel on $J$. Moving the finite alternating sum inside this common integral gives
\begin{equation}
\begin{aligned}
G_{ij}^{J,\delta}(p)
&=\sum_{\substack{S\subseteq J\\j\in S}}
(-1)^{|J|-|S|}x_{ij}^{S,\delta}(p_S)\\
&=\int_{(0,\infty)^{m\times J}}
\left(\sum_{\substack{S\subseteq J\\j\in S}}
(-1)^{|J|-|S|}x_{ij}^S(p_S+r_S)\right)
\kappa_{J,\delta}(r)\,dr\\
&=(G_{ij}^J\ast_{\mathrm R}\kappa_{J,\delta})(p)=0.
\end{aligned}
\label{eq:regularized-zero-interaction}
\end{equation}
The same compatibility applies to utilities, with the empty-set utility contributing zero. Since $(p_S)^{i,0}=(p^{i,0})_S$, the normalization terms combine in the same way, giving
$$
\begin{aligned}
V_i^{J,\delta}(p)
&=\sum_{S\subseteq J}(-1)^{|J|-|S|}u_i^{S,\delta}(p_S)\\
&=(V_i^J\ast_{\mathrm R}\kappa_{J,\delta})(p)
 -(V_i^J\ast_{\mathrm R}\kappa_{J,\delta})(p^{i,0})\\
&=0.
\end{aligned}
$$
Thus a single smooth truthful mechanism family has all interactions of order at least $m$ equal to zero on every nonnegative report, with the stated additive approximation error.
\end{proof}

\begin{proof}[Proof of Lemma~\ref{lem:bounded-interaction-replacement}]
Fix any $\delta>0$. Lemma~\ref{lem:padding-invariant-family} gives a padding-invariant truthful family with approximation guarantee $\Gamma$. Apply Lemma~\ref{lem:truthful-smoothing} to this family at the fixed scale $\delta$. The resulting allocation rules are feasible, have interaction order at most $m-1$, and satisfy the makespan bound \eqref{eq:regularized-performance}. This proves the lemma.
\end{proof}

When $\Gamma<(m+1)/2$, the proof of Theorem~\ref{thm:main} fixes an intermediate factor $\gamma>\Gamma$ and the bound $L=L(m,\gamma)$ supplied by Lemma~\ref{lem:bounded-interaction-lower-bound}, then chooses $\delta>0$ with $\Gamma mL^2\delta\le\gamma-\Gamma$. This absorbs the additive error on the hard instance obtained for the replacement mechanism. The earlier padding construction still uses compactness, but there is no additional limit as the smoothing scale tends to zero.

\begin{remark}[Why not apply left smoothing to the allocation rule?]
Direct left smoothing of $x$ may query reports $p-r$ with negative processing times, where the allocation rule is not defined. Extending $x$ by zero does not resolve this problem: at those reports, it gives $\sum_i x_{ij}=0$ instead of $1$, so averaging can violate complete allocation. For example, at $p=0$, every report sampled by left smoothing has negative coordinates, and the zero extension would yield a zero allocation. We therefore use zero extension and left smoothing only for the auxiliary interactions, and right smoothing to construct the actual mechanism.
\end{remark}

\begin{remark}[Measurability of the smoothing averages]
The smoothing integrals above implicitly assume joint measurability of the allocation rules and utilities. Even when the original mechanism is measurable, the compactness argument used for padding does not guarantee that its replacement remains measurable. Appendix~\ref{app:nonmeasurable-rules} removes this assumption by replacing each smoothing integral with a compatible positive linear averaging functional $I_D$, defined even for nonmeasurable bounded compactly supported integrands. These functionals retain the translation invariance and compatibility across job subsets needed in the proof. The appendix verifies that the smoothness, truthfulness, approximation, and vanishing conclusions remain valid under this substitution.
\end{remark}

\section{Conclusion}
\label{sec:conclusion}

Our proof develops job allocation interaction as a way to connect truthfulness with approximation lower bounds. We reduce truthful fractional mechanisms to allocation rules with bounded interaction and then derive the lower bound from bounded interaction alone. This separates the structural consequences of truthfulness from the construction of hard instances.

The interaction viewpoint may also be useful in other mechanism-design problems with additive costs or values. In such settings, the relation between allocations and utility derivatives suggests a possible route from incentive constraints to restrictions on interactions among items. Understanding when such restrictions can be obtained, and how they limit approximation, could extend this approach beyond scheduling.

\paragraph{AI disclosure.}
GPT-6 Astra (extra high) assisted with the proofs and writing of this paper. The authors developed the understanding of the problem presented in this paper, determined the messages to convey to readers, and selected, organized, and presented the key technical insights accordingly. The authors take full responsibility for the content and correctness of the paper.

\bibliographystyle{alpha}
\bibliography{references}

\clearpage
\appendix
\section{Removing the Measurability Assumption}
\label{app:nonmeasurable-rules}

If the allocation rules and utilities are not measurable, the smoothing averages in Section~\ref{sec:interaction-regularization} need not exist as ordinary integrals. In this case, we replace each such integral $\int_{\mathbb R^D}h(s)\,ds$ by a value $I_D(h)$, where $I_D$ is defined for every bounded compactly supported real function, whether measurable or not. The construction uses invariant means, a classical tool in functional analysis and the theory of amenable groups. Multiplication by the smoothing kernel puts each integrand in this class. We first state the properties that $I_D$ shares with ordinary integration, then explain why they suffice for the claims in the main proof. This removes the temporary assumption in footnote~\ref{fn:measurability}, including for the padding-invariant family constructed by compactness.

\subsection{Properties Shared with Ordinary Integration}

\begin{lemma}[Compatible positive integration]
\label{lem:compatible-positive-integration}
For every dimension $D$, there is a real-valued functional $I_D$ on all bounded compactly supported real functions on $\mathbb R^D$. These functionals can be chosen to satisfy the following properties:
\begin{enumerate}[label=(\arabic*)]
\item \emph{Linearity.} For all functions $h,g$ in this domain and all $a,b\in\mathbb R$,
$$
I_D(ah+bg)=aI_D(h)+bI_D(g)~.
$$
\item \emph{Positivity.} If $h\ge0$ pointwise, then $I_D(h)\ge0$. In particular, pointwise inequalities are preserved.
\item \emph{Translation invariance.} For every $a\in\mathbb R^D$,
$$
I_D\bigl(s\longmapsto h(s+a)\bigr)=I_D(h)~.
$$
\item \emph{Agreement with ordinary integration.} If $h$ is continuous and compactly supported, then $I_D(h)=\int_{\mathbb R^D}h(s)\,ds$. Also, $I_D(\mathbf1_B)=\operatorname{vol}(B)$ for every bounded box $B$, with any choice of endpoints.
\item \emph{Compatibility across coordinate subsets.} In an average with a normalized product kernel, deleting coordinates on which the averaged function does not depend gives the same average on the remaining coordinates, in their inherited order.
\end{enumerate}
\end{lemma}

These are the properties of integration used in our argument. In particular, a continuous nonnegative compactly supported kernel $\kappa$ of integral one satisfies $I_D(\kappa)=1$, so its weighted average preserves constants and pointwise bounds. Agreement with ordinary integration is asserted for continuous compactly supported functions and box indicators; for general integrands, the five properties above are what the proof requires.

\begin{proof}
Consider the positive unital linear functionals $\mathcal M$ on bounded real functions on the circle $\mathbb R/\mathbb Z$. They form a nonempty compact convex set in the product of the intervals $[-\|f\|_\infty,\|f\|_\infty]$: point evaluation gives an element, and linearity, positivity, and $\mathcal M(1)=1$ are closed conditions. Circle translations induce commuting continuous affine maps on this set. Lemma~\ref{lem:padding-common-fixed-point} gives a translation-invariant functional $\mathcal M$.

For continuous periodic $f$, the functions $K^{-1}\sum_{r=0}^{K-1}f(\cdot+r/K)$ converge uniformly to the integral of $f$ over one period. Their $\mathcal M$-values all equal $\mathcal M(f)$, so $\mathcal M$ agrees with that integral. For bounded compactly supported $f$ on $\mathbb R$, define
$$
I_1(f)=\mathcal M\left(s\longmapsto\sum_{r\in\mathbb Z}f(s+r)\right)~.
$$
The periodization is bounded because the sum has uniformly finitely many nonzero terms on a period. This defines a positive linear functional invariant under every real translation. For continuous $f$, its value is the usual integral. Continuous upper and lower approximations to interval indicators show that their values are the interval lengths, with any choice of endpoints.

Fix this same $I_1$ in every coordinate. Define $I_D$ by iterating it in a fixed coordinate order, integrating the last coordinate first. Each intermediate function is bounded and compactly supported: if the current function is bounded by $B$ and supported in a box, integration in one coordinate gives a bound of $B$ times that box's side length. Positivity and translation invariance persist under iteration. Agreement on continuous functions follows from ordinary iterated integration, and the box formula follows from the one-dimensional formula.

For report matrices, order coordinates by machine index and then by increasing job label. If a function depends only on a subset of the coordinates, an inactive coordinate occurs only in its normalized kernel factor. Its integration removes that factor with value one. The remaining order is the inherited order on the smaller report matrix, proving compatibility. No interchange of integrations of arbitrary nonmeasurable functions is required.
\end{proof}

\subsection{Why the Auxiliary-Function Claims Still Hold}

In place of the ordinary convolution, for a locally bounded function $f$ and a smooth compactly supported kernel $\kappa$ we now use
\begin{equation}
(f\ast_{\mathrm L}\kappa)(p)=I_D\bigl(s\longmapsto f(s)\kappa(p-s)\bigr)~.
\label{eq:nonmeasurable-smoothing}
\end{equation}
For the left-smoothed interactions in Section~\ref{sec:interaction-regularization}, we use this convolution with $f=V_i^J$ and a positive-side kernel $\kappa$.

\paragraph{Well-definedness and smoothness.}
For each $p$, the integrand in \eqref{eq:nonmeasurable-smoothing} is bounded and compactly supported, so $I_D$ assigns it a finite value. To justify differentiation, positivity and agreement on continuous functions give, for every continuous compactly supported $\varphi$,
$$
|I_D(f\varphi)|\le\sup_{\operatorname{supp}\varphi}|f|\int_{\mathbb R^D}|\varphi(s)|\,ds~.
$$
On a compact set of centers $p$, the kernel derivatives and their difference quotients have a common compact support. The kernel difference quotients converge uniformly there. The displayed estimate therefore allows differentiation under $I_D$ at every order and proves continuity of the resulting derivatives. Thus the auxiliary functions are smooth. This verifies the smoothness claim using uniform convergence, without invoking dominated convergence for $I_D$.

\paragraph{The derivative identity and conservation law.}
We also need the derivative of the smoothed utility interaction to equal the negative smoothed allocation interaction. The main-text proof uses integration by parts. After replacing the integrals by $I_D$, we prove the same identity using the pointwise incentive inequalities, positivity, and translation invariance.

Fix a job set $J$ with $n=|J|\ge2$, and write $D=mn$ and $N=2^{n-1}$. We use the zero extensions and bounds from Section~\ref{sec:interaction-regularization}.

\begin{lemma}[Smoothed derivative identity without measurability]
\label{lem:nonmeasurable-smoothed-derivative}
For the zero-extended interactions of the padding-invariant family and every nonnegative smooth compactly supported kernel $\kappa$, the convolution defined in \eqref{eq:nonmeasurable-smoothing} satisfies
$$
\frac{\partial}{\partial p_{ij}}(V_i^J\ast_{\mathrm L}\kappa)
=-(G_{ij}^J\ast_{\mathrm L}\kappa),
\qquad
\sum_{i=1}^m\frac{\partial}{\partial p_{ij}}(V_i^J\ast_{\mathrm L}\kappa)=0~.
$$
Thus Lemma~\ref{lem:smoothed-utility-derivative} remains valid with $I_D$ in place of integration.
\end{lemma}

\begin{proof}
Fix $e=e_{ij}$ and $h>0$. For $S\subseteq J$ containing $j$, let $B_S(s)=x_{ij}^S(s_S)$ for $s\in[0,\infty)^D$, and let $B_S(s)=0$ otherwise. Set
$$
E_h(s)=\frac{V_i^J(s+he)-V_i^J(s)}h+G_{ij}^J(s)~.
$$
For $s\in[0,\infty)^D$, including boundary reports, the difference-quotient bounds derived from \eqref{eq:normalized-ic} bound each individual utility residual between zero and $B_S(s)-B_S(s+he)$. Taking the absolute value after the signed sum therefore bounds $|E_h(s)|$ by the sum of these nonnegative decreases. This applies the incentive inequalities to the individual utilities before forming their alternating sum.

If another coordinate is negative, or if $s_{ij}<-h$, the relevant extended functions vanish at both reports. In the remaining case, all other coordinates are nonnegative and $-h\le s_{ij}<0$. Here $V_i^J(s)=G_{ij}^J(s)=0$, and the zero boundary condition and the Lipschitz bound give $|V_i^J(s+he)|\le Nh$, so $|E_h(s)|\le N$. Since the sum of the marginal differences is at least $-N$, we have everywhere
$$
|E_h(s)|\le
\sum_{\substack{S\subseteq J\\j\in S}}
\bigl(B_S(s)-B_S(s+he)\bigr)
+2N\mathbf1_{[-h,0)}(s_{ij})~.
$$
Multiply by $\kappa(p-s)$ and apply $I_D$. Translation invariance and $0\le B_S\le1$ bound the absolute value of each integrated difference by $\|\kappa(\cdot+he)-\kappa\|_{L^1}=O(h)$. In a fixed box containing the kernel support, the strip term also has integral $O(h)$ by the box formula. Thus $I_D[E_h(\cdot)\kappa(p-\cdot)]\to0$.

Translation invariance identifies the utility term in this expression with the difference quotient of $V_i^J\ast_{\mathrm L}\kappa$ at $p$. Smoothness then gives \eqref{eq:smoothed-utility-derivative}. Summing over machines and using \eqref{eq:interaction-conservation} proves the second identity in \eqref{eq:smoothed-utility-derivative}.
\end{proof}

\paragraph{Zero boundaries and vanishing interactions.}
For a nonnegative smooth kernel $\kappa$ of integral one with compact support in $(0,\infty)^D$, the zero boundary argument is pointwise: when $p_{ij}=0$, every nonzero kernel weight samples a negative own coordinate, where the extended $V_i^J$ is zero. Applying $I_D$ to this zero integrand still gives zero. Moreover, Lemma~\ref{lem:nonmeasurable-smoothed-derivative}, positivity, and $|G_{ij}^J|\le N$ give $|\partial(V_i^J\ast_{\mathrm L}\kappa)/\partial p_{ij}|\le N$.

Together with smoothness and the conservation law, these verify all four conditions of Lemma~\ref{lem:laplace-annihilation}. Thus the proof of Lemma~\ref{lem:vanishing-auxiliary-functions} still applies, and gives $V_i^J\ast_{\mathrm L}\kappa=0$ when $|J|\ge m$. The derivative identity then gives $G_{ij}^J\ast_{\mathrm L}\kappa=0$ as well. At this stage the auxiliary functions are smooth ordinary functions, so the Laplace transforms in Lemma~\ref{lem:laplace-annihilation} are ordinary integrals. No replacement of the Laplace-transform argument is needed.

\subsection{Why the Mechanism Claims Still Hold}

For the actual mechanism, interpret right smoothing by
$$
\begin{aligned}
(h\ast_{\mathrm R}\kappa_{S,\delta})(p)
&=I_{m|S|}\left(s\longmapsto h(s)\kappa_{S,\delta}(s-p)\right)\\
&=I_{m|S|}\left(r\longmapsto h(p+r)\kappa_{S,\delta}(r)\right).
\end{aligned}
$$
Here $h$ is extended by zero, and the second equality uses translation invariance. For $p\ge0$, the integrand in the second line is supported on strictly positive perturbations $r$. Use the coordinate order from Lemma~\ref{lem:compatible-positive-integration} for every job set, and interpret the convolutions in \eqref{eq:regularized-mechanism} using these functionals. We check each conclusion of Lemma~\ref{lem:truthful-smoothing} under this substitution.

\begin{enumerate}[label=(\arabic*)]
\item \emph{Smoothness and feasibility.} The integrands are bounded on the kernel support, so the allocation and utility averages are finite. Each unnormalized right average pairs $h(s)$ with the smooth compactly supported kernel $\kappa_{S,\delta}(s-p)$. The smoothness argument above therefore applies, including at boundary reports. Positivity and $I_{m|S|}(\kappa_{S,\delta})=1$ preserve $x_{ij}^{S,\delta}\ge0$ and $\sum_i x_{ij}^{S,\delta}=1$.
\item \emph{Truthfulness.} For every fixed perturbation $r$, the original incentive inequality holds pointwise. Linearity and positivity allow us to average this inequality exactly as in the main proof. Subtracting the reference utility preserves the incentive inequality and enforces zero-row normalization. The resulting allocations and utilities therefore admit the same finite implementing payments.
\item \emph{Approximation guarantee.} Each perturbation $r$ with $\kappa_{S,\delta}(r)\ne0$ satisfies $0<r_{ij}<\delta$. The pointwise load bounds in the main proof, together with $\operatorname{OPT}(p+r)\le\operatorname{OPT}(p)+|S|\delta$, remain valid after applying the positive normalized average. Hence the makespan is at most $\Gamma(\operatorname{OPT}(p)+|S|\delta)$.
\item \emph{Vanishing interactions.} Linearity allows the finite alternating sums defining interactions to pass through the average, and compatibility across job subsets identifies all terms with averages from the same mechanism family. The symmetry identity $\kappa_{J,\delta}(s-p)=\kappa_{J,\delta}(p+\delta\mathbf1-s)$ gives the same correspondence between right averages and shifted auxiliary functions as in \eqref{eq:smoothing-translation}. Thus the new utility interaction is the difference of two zero auxiliary-function values, and the averaged allocation interaction is also zero. This proves \eqref{eq:regularized-zero-interaction} at every nonnegative report; every report queried by the right average is strictly positive.
\end{enumerate}

Consequently, replacing the smoothing integrals by $I_D$ preserves all the claims needed for Lemma~\ref{lem:bounded-interaction-replacement}. For every $\delta>0$, we obtain a truthful mechanism family of interaction order at most $m-1$ satisfying
$$
C(p,x^{S,\delta}(p))\le\Gamma\bigl(\operatorname{OPT}(p)+|S|\delta\bigr)~.
$$
No measurability assumption is needed on either the original mechanism or the padding-invariant family.

\end{document}